\documentclass[twocolumn,showpacs,preprintnumbers,amsmath,amssymb,aps]{revtex4}
\usepackage{latexsym}
\usepackage{amsfonts}
\usepackage{amsmath}
\usepackage{bm}
\usepackage{amssymb}
\usepackage{epsfig}
\begin{document}

%\preprint{APS/123-QED}

\title{Phase space structures and ionization dynamics of hydrogen atom
in elliptically polarized microwaves}

\author{E. Shchekinova$^1$\footnote{elena@cns.physics.gatech.edu}, C. Chandre$^2$ and T. Uzer$^1$}

\affiliation{$^1$Center for Nonlinear Science, School of Physics,
Georgia Institute of Technology, Atlanta, Georgia 30332-0430, U.S.A.\\
$^2$Centre de Physique Th\'eorique%
\footnote{Unit\'{e} Mixte de Recherche (UMR 6207) du CNRS, et des universit\'{e}s
Aix-Marseille I, Aix-Marseille II et du Sud Toulon-Var. Laboratoire
affili\'{e} \`{a} la FRUMAM (FR 2291).%
} - CNRS, Luminy - Case 907,
13288 Marseille cedex 09, France}

\date{\today}% It is always \today, today,

\begin{abstract}
 The multiphoton ionization of hydrogen atoms in a strong
elliptically polarized microwave field exhibits complex
features that are not observed for ionization in
circular and linear polarized fields. Experimental data reveal high sensitivity of ionization dynamics
to the small changes of the field polarization. The
multidimensional nature of the problem makes widely used
diagnostics of dynamics, such as Poincar\'{e} surfaces of section,
impractical. We analyze the phase space dynamics using finite time
stability analysis rendered by the fast Lyapunov Indicators
technique. The concept of zero--velocity surface is used to
initialize the calculations and visualize the dynamics. Our
analysis provides stability maps calculated for the
initial energy at the maximum and below the saddle of the
zero--velocity surface. We estimate qualitatively the dependence of
ionization thresholds on the parameters of the applied field, such
as polarization and scaled amplitude.
%The non--monotonic increase of ionization probabilities
%is linked to the stabilization and collapse of the resonant tori
%structures near the maximum of zero--velocity surface. The
%application of the fast Lyapunov Indicators analysis reveals
%significant variations in the phase space stability structure and
%ionization dynamics that happen with the change of fields
%polarization degree.
\end{abstract}


%\tableofcontents

\maketitle

\section{Introduction}
\label{sec1}

 The multiphoton ionization of the hydrogen atom in a strong microwave field is an example of
a low--dimensional quantum system that manifests behavior typical
of classical nonlinear systems, such as chaos and high sensitivity to field parameters. Ever since the first experiments on the
ionization of hydrogen atoms in a strong microwave field
\cite{bayfield74} this system has been studied extensively for
various regimes of the applied field: From high-- to
low--frequency regimes, and from linear to circular polarization
of the field \cite{koch95}. The ionization dynamics turns out to
be strongly dependent on the parameters of the applied field, such
as frequency, field amplitude, the shape of the field envelope,
and the choice of initial ensemble of states subjected to the
field \cite{beller97,brun97}.

  Experiments \cite{beller97,beller96} on the ionization
of hydrogen atom in strong elliptically polarized (EP) microwaves
have demonstrated the dynamics in this case to be quite
complex, exposing new effects that are absent from the circularly
polarized (CP) and linearly polarized (LP) field limits. The
ionization dynamics was observed to be extremely sensitive to the
polarization degree of the electric field. Moreover, three-dimensional classical simulations of  the experiment,
reproduce the experimental results very well
\cite{beller96,rich97}. It was shown in \cite{beller97} that the
ionization yields, deduced experimentally and
numerically, interpolate unevenly between LP and CP limits and for
the EP fields follow a non--monotonic rise with the increase of
the scaled field amplitude. By viewing the system in the
frame rotating with the time--dependent angular velocity of the EP
field the authors in \cite{oks83,oks99} used a quantum--mechanical
approach to show that the problem can be reduced to a Rydberg atom
subjected to two effective static fields, magnetic and electric.
The sensitivity of the ionization to the ellipticity
degree observed experimentally was ascribed to dependence of the
effective magnetic interaction on the ellipticity degree. While
the use of the quantum--mechanical approach for hydrogen in an EP
field reproduces the experimental trends, a connection with the
classical approach, which has been traditional in this subject, is desirable.
The question still remains open: Why are the current
experimental data on ionization yield curves of hydrogen atom in
EP microwave fields in good agreement with the
corresponding classical simulations \cite{beller97,fu90}?

Most classical treatments in the
literature use some simplifications of dynamics based on the
existing symmetries or reduction of the full three-dimensional
dynamics of the system to a lower--dimensional one
\cite{rich97,griff92,kappertz93}. To give an example, the
adiabatic approximation used by Griffith and Farrelly
\cite{griff92} to study three-dimensional hydrogen atom in EP field
reduces the dimensionality of the effective phase space and limits the
dynamics to that of the circular Rydberg states. The averaging
that is performed in \cite{rich97} is valid only in
the low--frequency regime (the driving frequency of the field is
much smaller than the unperturbed Kepler frequency of the
problem). The presence of a
time--dependent term in the rotating-frame Hamiltonian is another challenge that the EP field brings. In
addition, the absence of conserved quantities, such as energy and
angular momentum integrals, prevents one from reducing the dimensionality of the dynamics. For the LP and CP problems such a
reduction is possible due to existence of additional constants of
motion. In contrast, for the EP field problem the energy and
angular momentum are not conserved. In this paper we discuss the
possibility to view a full--dimensional system dynamics by means
of fast Lyapunov Indicator (FLI) fields of stability
\cite{froe97}. The FLI method is independent of the
dimensionality of the system. It provides pictures of global stability
structures for any given subspace of the phase space of the
system \cite{shch04}. Besides, the method is robust because it is applicable even in regimes which cannot be studied by any other existing analytical methods.

The complex features of ionization dynamics of hydrogen atom in EP
microwave field reported in recent experiments \cite{beller97} are
due to the higher dimensionality of the EP problem as
opposed to the CP and LP problems. Classically, for systems of
three or higher degrees of freedom the phase space dynamics is
known to possess much richer features than the one for the lower--
dimensional systems. The principal feature that distinguishes the
phase space dynamics of Hamiltonian system of dimension three and
higher is the possibility of a well--known phenomenon of Arnold
diffusion (diffusion along the resonances) \cite{arnold89}. This phenomenon is essential for examining all
the possible scenarios that lead to the classical ionization for
hydrogen subjected to the electromagnetic field.

The relevance of
the phase space dynamics to the classical ionization of hydrogen
in non--stationary fields has been noted by many authors
\cite{koch95,milcz96,farrelly95,chand02}. In short, invariant tori
can prevent chaotic diffusion of trajectories and, therefore, can
hinder them from ionizing. The onset of stochastic
ionization for the Hamiltonian systems has been traditionally
investigated by means of the resonance overlap criterion proposed by
Chirikov \cite{chirik79}. It was applied to the classical
ionization dynamics for the hydrogen atom driven by LP and CP
field in
\cite{koch95,chand02,casati87,jensen84,howard92,blumel94}. The set
of action--angle variables required for this analysis is only well--defined in the limit of
vanishing field. For the classical system of hydrogen atom
subjected to a strong EP field the application of action-angle
variables in an overlap criterion is not practical. Instead, we propose to study the onset of
stochasticity in the phase space of the system by means of fast
Lyapunov Indicator (FLI) stability analysis \cite{froe97}. By evaluating
the indicator of chaoticity for each integrated trajectory from an
ensemble of trajectories we obtain the underlying resonant structures
for any given subspace.

The experimental data \cite{beller97} do not provide much
information about the character of the states that determine the
observed ionization threshold. In fact, most of the initial states
used in the experiment are switched by the turn--on of the field
to different locations in the phase space \cite{kappertz93,farrelly95}. Some
of the states that remain bounded after the rise of the field
pulse are closer to stochastic ionization than the other states. The character of the states more favorable to
ionization has been a matter of debate in the theory community. Investigations on the ionization by the CP field in
high--frequency regime were carried out by Howard \cite{howard92}
and  by Sacha and Zakrzewski \cite{sacha97}. Using essentially the
same technique, the Chirikov overlap criterion, these two groups
arrived at different conclusions regarding to the type of the
orbits that ionize first under the application of the field.
Howard claimed that most orbits become elongated before
they ionize, hence that the high eccentricity orbits are more
prone to ionization. However, results in \cite{sacha97}
contradict this point of view and the authors show that ionization
occurs first for the medium eccentricity orbits. In Refs.
\cite{kappertz93,farrelly95} it was pointed out that the ionization mechanism
strongly depends on the region of phase space that the
original ensemble will be switched to by the application of the
field. Generally speaking, orbits with different eccentricities
can be switched to different regions of the phase space and to
different energies in the rotating frame of the CP field problem.
In this paper we describe in details the character of the states
that undergo ionization and determine the behavior of ionization
yields for the low amplitudes of the field. The principal
difficulty is the direct comparison of our results with
experiments \cite{beller97}. It is essential to point
out, that our qualitative analysis is performed for ensembles
of states with the same initial energy and different values of
classical action and angular momentum, as opposed to the ensemble
of states used in the experiment. Our purpose in this choice is
to relate the underlying phase space structures to
the ionization dynamics of the system.

The recent experimental success in creating field-maintained non-spreading electronic wave packets \cite{maeda04} has renewed the discussion on synthesizing coherent
states launched from the Lagrange equilibria of the effective
potential \cite{bialyn94, brunello96, Lee97}. It was argued in Ref.
\cite{Lee97} that the size of the stability region surrounding the
Lagrange equilibrium points is important in maintaining stable
non--spreading three-dimensional coherent quantum states at those points. The
addition of a magnetic field to the CP problem was shown to play
an important role in the stabilization of the equilibria and
enlarge the critical region of stability in the parameter space
defined by the Jacobi constant, and the amplitudes of the electric and magnetic
fields. We extend these findings to the EP problem by using fast
Lyapunov Indicator stability technique. We show that the stability of the Lagrange equilibria can be controlled
by manipulating the field polarization in addition
to the Jacobi constant, and the amplitudes of the electric and magnetic
fields.

Much of our understanding of complex details of the classical
ionization mechanism depends on a thorough knowledge of the
multidimensional dynamics of the system. Therefore, we will apply the FLI method so as to obtain some insight into the stability of the phase-space structures of the system.

 The paper is organized in five sections. In Sec.~\ref{sec2} we
introduce the Hamiltonian and describe the concept of
time--dependent zero--velocity surface (ZVS) in the frame rotating
with the frequency of the applied field. A brief description of
the method of Fast Lyapunov indicators (FLI) is given in
Sec.~\ref{sec3}. We discuss the results of the stability analysis
rendered by FLI method in Sec.~\ref{sec4}. By mapping out the
value of the FLI for each trajectory from the configuration space
we obtain the FLI stability plots. First, the structures of the
FLI stability plots are described for the well--studied cases of
LP and CP electric fields and zero magnetic field. The energy of
the initial states is equal to the maximum of the ZVS. In the case of
the CP problem the FLI plot matches perfectly the Poincar\'{e}
surface structure. Secondly, we perform the FLI analysis for more
complicated problems of hydrogen in EP fields (the magnetic field
is again equal to zero). The ionization dynamics for the EP problem is
analogous to the LP and CP problems: two distinct sets of bounded
orbits located around the center and near the maximum of the ZVS
are identified. With the increase of the amplitude of the electric
field the orbits from the latter set become chaotic and ionize.
The size of the stability zone around the ZVS maximum is used to
estimate qualitatively the behavior of the ionization probabilities
versus scaled amplitude of the field. Then, the FLI stability analysis
is applied to the ensemble of initial states with the energy much
below the saddle of the ZVS. Section \ref{sec4} concludes with
the detailed description of two ensembles of initial states
involved in our calculations. A summary of our results and
conclusions are presented in Sec.~\ref{sec5}.

\section{Hamiltonian}
\label{sec2}

The Hamiltonian for a hydrogen atom subjected to an EP
electric field (of magnitude $F$ and microwave frequency $\omega$)
simultaneously with the static magnetic field ${\bf B}=B\hat{{\bf z}}$ applied
perpendicular to the plane of polarization is, in atomic units ($a_0=\hbar=e=\mu=1$) and assuming infinite
nuclear mass, as follows:
\begin{eqnarray}
 H& = &
 \frac{1}{2}(p_{x}^2+p_{y}^2+p_z^2)-\frac{1}{2}B(x p_{y}-y p_{x})\nonumber\\
 & &-\frac{1}{r}+ \frac{1}{8}B^2(x^2+y^2)\nonumber\\
 & &+ F(x\cos \omega t+\alpha y\sin \omega t)~,
\end{eqnarray}
where $r=\sqrt{x^2+y^2+z^2}$ and $\alpha\in[0,1]$ is a polarization of the electric field~:
$\alpha=1$ for the circularly polarized field, and $\alpha=0$ for
the linearly polarized field. In this paper we study the planar
classical model of hydrogen restricted to the $x-y$ polarization
plane. Most of the previous classical and quantum calculations
were performed in the planar limit
\cite{brun97,griff92,kappertz93,farrelly95,Lee97}. The planar
limit of $3$D system is a reliable approximation to the actual
three--dimensional dynamics for the orbits with initial conditions
initiated on the $z=0,p_z=0$ subspace. This subspace is invariant.
Moreover the classical phase space dynamics for the planar limit
of the hydrogen atom inside CP and LP fields can be studied by
means of Poincar\'{e} surfaces of section, which in this case is
two--dimensional. Since the applied electric field is polarized in
the $x-y$ plane the ionization of hydrogen is expected to occur
along any direction on the plane as in the Stark effect.

% We consider the case of
%$\omega=\omega_K$, where $\omega_K$ is a Kepler frequency of the
%unperturbed problem.
Insight into the dynamics can be obtained by considering
the system in the rotating frame precessing with the frequency of
the field $\omega$. By introducing the canonical transformation
\begin{eqnarray}
S(\bar{x},\bar{y},p_x,p_y)&=&-(\bar{x}\cos\omega
t-\bar{y}\sin\omega t)p_x\nonumber\\& &-(\bar{x}\sin\omega
t+\bar{y} \cos\omega t)p_y,
\end{eqnarray}

one obtains the Hamiltonian in the rotating frame (for
convenience we omit the bars in the following expression and
thereafter):
\begin{eqnarray}
H &=& \frac{1}{2}(p_{x}^2+p_{y}^2)-(1+\frac{1}{2}B)(x
p_{y}-y p_{x})\nonumber\\
& &-\frac{1}{r}\nonumber+ \frac{1}{8}B^2(x^2+y^2)+F x\nonumber\\
& &+ F(\alpha-1)(x\sin^{2}\omega t+y\sin \omega t\cos \omega t)~.
\end{eqnarray}
It is more convenient to work with scaled frequencies and field
strengths. We introduce the scaling of time, coordinates, momenta,
and field amplitudes as follows: $t^{'}=\omega t,
r^{'}=\omega^{2/3}r, p^{'}=\omega^{-1/3}p, K=\omega^{-2/3}H,
F^{'}=\omega^{-4/3}F, B^{'}=\omega^{-1}B$. After dropping the
primes, the scaled Hamiltonian becomes:
\begin{eqnarray}
K&=& \frac{1}{2}(p_{x}^2+p_{y}^2)-(1+\frac{1}{2}B)(x
p_{y}-y p_{x})\nonumber\\
& &-\frac{1}{r}\nonumber+ \frac{1}{8}B^2(x^2+y^2)+F x\nonumber\\
& &+ F(\alpha-1)(x\sin^{2} t+y\sin t\cos t)~.
\end{eqnarray}

For the CP field ($\alpha=1$) the Hamiltonian is
 time--independent and the system has
two effective degrees of freedom. The Jacobi constant introduced
in celestial mechanics \cite{hill78,szebehely67} is equal to the
energy of the system in the rotating frame. For the EP field
($0< \alpha < 1$) the Hamiltonian is a time--dependent quantity.
For the EP problem due to the absence of integrals of motion the
system cannot be reduced to the two--dimensional one as was done for the
CP problem. Instead, the time--dependent Hamiltonian in $(4)$ can
be made autonomous by introducing canonical transformation
$S(x,y,p_x,p_y,t,w)=w K(x,y,p_x,p_y,t)$ where $w=t$ is a
generalized coordinate. The corresponding generalized momentum is
defined as  $p_w=-K(t)$. The transformation yields the expression
for the effective autonomous Hamiltonian:
\begin{eqnarray}
H_{eff} & = & \frac{1}{2}(p_{x}^2+p_{y}^2)+p_w -(1+\frac{1}{2}B)(x
p_y-y p_x)\nonumber\\& &-\frac{1}{r}
+\frac{1}{8}B^2(x^2+y^2)\nonumber\\& &+F(\alpha-1)(x\sin^2 w+y\cos
w \sin w)~.
\end{eqnarray}
Equation $(4)$ describes an autonomous
Hamiltonian system with three degrees of freedom . In practice, the Hamiltonian is regularized by
changing coordinates to the set of semi--parabolic coordinates.
This substitution is used in order to avoid the singularity near
the origin $(x,y)=(0,0)$ \cite{chand02}.

\begin{figure}
 \begin{center}
 \includegraphics[width=5.cm]{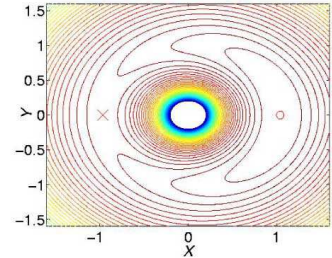}
 \caption {\label{fig:fig1} (Color online) Zero velocity surface contour plot at
 time $t=0$ for the hydrogen in the EP microwave electric field.
 $F=0.117, B=0$ (in scaled units). The circle (cross) indicates the location of the
 maximum (saddle) point of the ZVS.}
 \end{center}
\end{figure}

 An additional complexity in the problem comes from the the velocity--dependent Coriolis term proportional to $xp_y-yp_x $
in the expression for Hamiltonian $(5)$. In its presence, the Hamiltonian can
not be split into a positive definite kinetic term depending on
momenta alone and potential energy term depending exclusively on
positions. Because of the Coriolis term, the
potential energy surface cannot be used for understanding the
stability of equilibria. Instead, the concept of zero velocity
surface, which constitutes an effective potential, is adapted from
celestial mechanics \cite{hill78,szebehely67}. To define
the ZVS we express Hamiltonian
$(4)$ in terms of velocities and positions~:
\begin{eqnarray}
K & = & \frac{1}{2}(\dot x^2+\dot
y^2)-\frac{(1+B)}{2}(x^2+y^2)-\frac{1}{r} + F x\nonumber
\\& &+F(\alpha-1)(x\sin^2 t+y\cos t\sin t)~.
\end{eqnarray}
Setting velocities in the above expression to zero we arrive at the
following form of the effective time--dependent potential surface:
\begin{eqnarray}
V(x,y,t) & = & -\frac{1}{r}-\frac{(1+B)}{2}(x^2+y^2)
 + F x\nonumber \\& &+F(\alpha-1)(x\sin^2 t+y\cos t\sin t)~.
\end{eqnarray}
For the CP case ($\alpha=1$) the ZVS is time--independent in the
rotating frame. There are two equilibria lying on the $x$--axis
$(y=0)$ corresponding to the maximum $(+)$ and saddle point $(-)$
of the surface:
\begin{equation}
-(1+B)x^3_{\pm}+F[1+(\alpha-1)\sin^2 t]x^2_{\pm} \pm 1=0~.
\end{equation}
The expression for the maximum $K_{max}$ and the saddle $K_{sad}$
of the ZVS results:
\begin{eqnarray}
K_{max(sad)}& =&
-\frac{1}{|x_{\pm}|}-\frac{(1+B)}{2}x_{\pm}^2\nonumber\\
 & &+ F(1+(\alpha-1)\sin^2 t) x_{\pm}~.
\end{eqnarray}
Fig.~\ref{fig:fig1} shows the zero--velocity contour plot and the
location of the equilibria $x_{\pm}(0)$ at time
$t=0$.

 Unlike the maxima of the potential surface that are
always unstable, the maxima of the ZVS need not to be. In fact,
linear stability analysis carried out in the vicinity of
equilibria in Ref.~\cite{Lee97} demonstrates that in the CP limit
the stability of equilibria of the ZVS depends on the parameters
of the field such as a scaled amplitudes of the electric and
magnetic field. The equilibria for the CP problem play the same
role as the Lagrange points $L_{4}$ and $L_5$ for the restricted
three--body problem \cite{hill78,szebehely67}.

Equation $(7)$ shows that the ZVS of the EP problem is
not constant and oscillates around the origin with the frequency
of the applied field. The linear stability analysis of equilibria
performed in \cite{Lee97} for CP problem is not tractable for the
EP problem. We study local dynamics around the Lagrange maximum by
applying linear stability analysis rendered by the fast Lyapunov
indicator (FLI) method.

\section{Description of the fast Lyapunov indicator method}
\label{sec3}

To characterize the multidimensional phase space
structures a method is needed that provides a clear
representation of the chaotic and regular regions in the phase
space, similarly to the pictures obtained by Poincar\'{e}
sections. However for the three-- and higher--dimensional systems
the construction and visualization of Poincar\'{e} sections might be
very complicated. An alternative is to define an indicator of chaoticity
for each trajectory from a given subspace of phase space. Various
diagnostics for different types of dynamical behavior are present
in the literature; for instance, indicators of chaoticity are the
characteristic Lyapunov exponent \cite{lich83}, the strength of
diffusion of the instantaneous frequency in the frequency space
\cite{laskar93}, the measure of complexity of the Fourier spectrum
\cite{powell79}, and other available indicators that distinguish
chaotic and regular dynamics \cite{cincotta00,cipriani98}. As a
short--term dynamical diagnostic a method of Fast Lyapunov
indicators (FLI) was proposed in Ref.~\cite{froe97} to study weak
chaos in high--dimensional systems. The main idea of the method is
to study the evolution of the tangent vectors computed along a
given trajectory of the system. This method has been applied to
search for phase space structures in different types of problems:
coupled standard maps \cite{guzz02}, celestial mechanics problems
\cite{astakh04}, and vibrational dynamics of polyatomic molecules
\cite{shch04}. In this paper we apply this method to the atomic
system. Below we present a brief description of the method (for
more details, see Ref.~\cite{guzz02}).

To obtain a numerical estimate on the growth of tangent vectors
along the flow the equations of motion are integrated together
with the corresponding equations for Jacobian matrix, i.e., we
integrate the flow given by $f(X)$ and the tangent flow:
\begin{equation}
\frac{d}{dt} X = f(X)~, \qquad \qquad
 \frac{d\mathcal{J}}{dt} = \mathcal{D} f(X) \mathcal{J}~,
\end{equation}

where $X=(x_1,x_2,\ldots,x_n)$ is a $n$--dimensional vector in the
phase space, $\mathcal{J}$ is $n\times n$ Jacobian matrix of the
flow, $\mathcal{D} f(X)$ is the matrix  of partial derivatives
$\{\partial{f}_i/\partial{x}_j\}_{i,j=1,n}$ of the flow, and
${\mathcal J}(0)$ is the identity matrix. The columns of Jacobian matrix are the
tangent vectors of the flow $\{{\bf v}_j\}_{j=1,n}$. Given initial
tangent vector ${\bf v}_0$ for each initial condition ${\bf x}_0$
from the phase space of the system the fast Lyapunov indicator
(FLI) is defined at time $t$ as follows:
\begin{equation}
\phi(t;{\bf x}_0)=\max_{0\leq t'\leq t}\log \Vert {\bf v}(t';{\bf
x}_0)\Vert~,
\end{equation}
where ${\bf v}(t';{\bf x}_0)$ is the tangent vector at time $t'$.
For any initial choice of the tangent vector ${\bf v_0}$ it will
eventually converge to the direction of the most unstable
eigenvector of the Jacobian matrix $\mathcal{J}$. Hence,
characteristic dynamics can be observed by integrating Eqs.~$(10)$
for one of the tangent vectors from the set of all the tangent
vectors in $\mathcal{J}$.
\begin{figure}
 \begin{center}
 \includegraphics[width=5.cm]{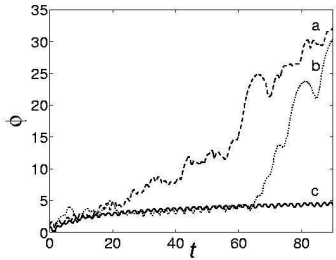}
 \caption {\label{fig:fig2} FLI curves for chaotic and regular
 trajectories of the hydrogen atom in a CP microwave field.
 $F=0.117, K_{max}=-1.3807, B=0$. Initial coordinates
$(x,y)$ for the corresponding
 trajectories are: a) (1.5, 0.01), b) (0.9,0.01), and c) (0.8, 0.01). The values of the initial
 momenta are chosen as described in Sec.~\ref{sec:4.1.1}. The time $t$ is scaled and
 dimensionless. }
\end{center}
\end{figure}

The Lyapunov indicator defined in Eq.~$(11)$ serves as an
efficient diagnostic of chaoticity for any analyzed trajectory
with initial condition ${\bf x}_0$. The FLI procedure supplies
information about the dynamical behavior of a trajectory, analogously to the method of
the characteristic Lyapunov exponents. The
tangent vector growth evaluated by integration of Eq.~$(10)$ is
proportional to the growth of the distance between two neighboring
trajectories in phase space. In fact, the evolution of the tangent
vector computed for chaotic trajectories obeys an overall
exponential law. Therefore, the FLI increases linearly in time. At
the same time, the tangent vector growth for regular trajectories
follows approximately a global linear behavior. Thus the FLI
increases logarithmically in time. Already after a short interval
of time it is possible to make a clear distinction between
different dynamical behaviors of trajectories by looking at the
maximum value attained by the FLI. To provide an example, in
Fig.~\ref{fig:fig2} the FLI evolution in time is shown for two
chaotic and one regular trajectories from the phase space of the
system defined by Hamiltonian $(5)$ . Already
 after a time $t=100$ two different types of dynamical behavior can be clearly identified.
 Meanwhile the curves $(a)$ and $(b)$ evolve almost linearly,
the curve $(c)$ follows logarithmic growth with time. Note that
$(b)$ and $(c)$ are launched very close and hence the FLI are very
close for some time before they are distinguished since $(b)$ is
chaotic and $(c)$ is regular (by inspections of their Poincar\'{e}
sections)

The main advantage of FLI technique over other available methods
is its power to discriminate regular from chaotic motions
over a relatively short time period.

\section{Numerical results}
\label{sec4}
\subsection{Ionization dynamics at the maximum of
zero--velocity surface}
\label{sec:4.1}

\subsubsection{Choice of initial conditions}
\label{sec:4.1.1}

We analyze the ionization dynamics of hydrogen atom in EP field by
applying FLI stability analysis for trajectories from a mesh of
initial conditions from the $x-y$ plane. For the CP problem we
consider the subspace of initial conditions defined in
Ref.~\cite{Lee97}. Namely, the initial energy is chosen to be
equal to the maximum of the ZVS: $K(0)=K_{max}$ as defined in
Eq.~$(9)$. The initial momenta $p_x, p_y$ are chosen to satisfy
the relation $p_x x+p_y y=0$.
%By substituting
%$p_x=-p_y\frac{y}{x}$ into the expression for the effective
%Hamiltonian $(5)$ for $\alpha=1$ one gets a quadratic equation for
%the momentum $p_y$ at initial time $t=0$:
%\begin{eqnarray}
%\frac{1}{2}\left(1+ \frac{y^2}{x^2}\right)p_y^2&
%&-(1+\frac{1}{2}B)\left(\frac{y^2}{x}+x\right)p_y\nonumber\\&
%&-\frac{1}{r} +\frac{1}{8}B^2\left(x^2+y^2\right)-K(0)=0~,
%\end{eqnarray}
%where $x\neq 0$. The singularity around $x=0$ is avoided by using
%semi--parabolic coordinates.

For the CP problem the above defined subspace of initial
conditions is two--dimensional and coincides with the
two--dimensional Poincar\'{e} section. Similarly to the CP problem
the dynamics for the EP problem is visualized on the $x-y$ plane.
For the EP problem the effective Hamiltonian of the system has
three degrees of freedom. In order to reduce the configuration
space of the system to two--dimensional in five--dimensional
energy subspace one needs to specify three initial conditions for
coordinates and momenta. One condition is the same as for the CP
problem. Additional conditions are defined by taking generalized
coordinate $w=0$ and initial generalized momentum $p_w=-K(0)$.

\subsubsection{FLI analysis for CP and LP fields}
\label{sec:4.1.2}

\begin{figure}
 \begin{center}
 \includegraphics[width=5.cm,angle=-90]{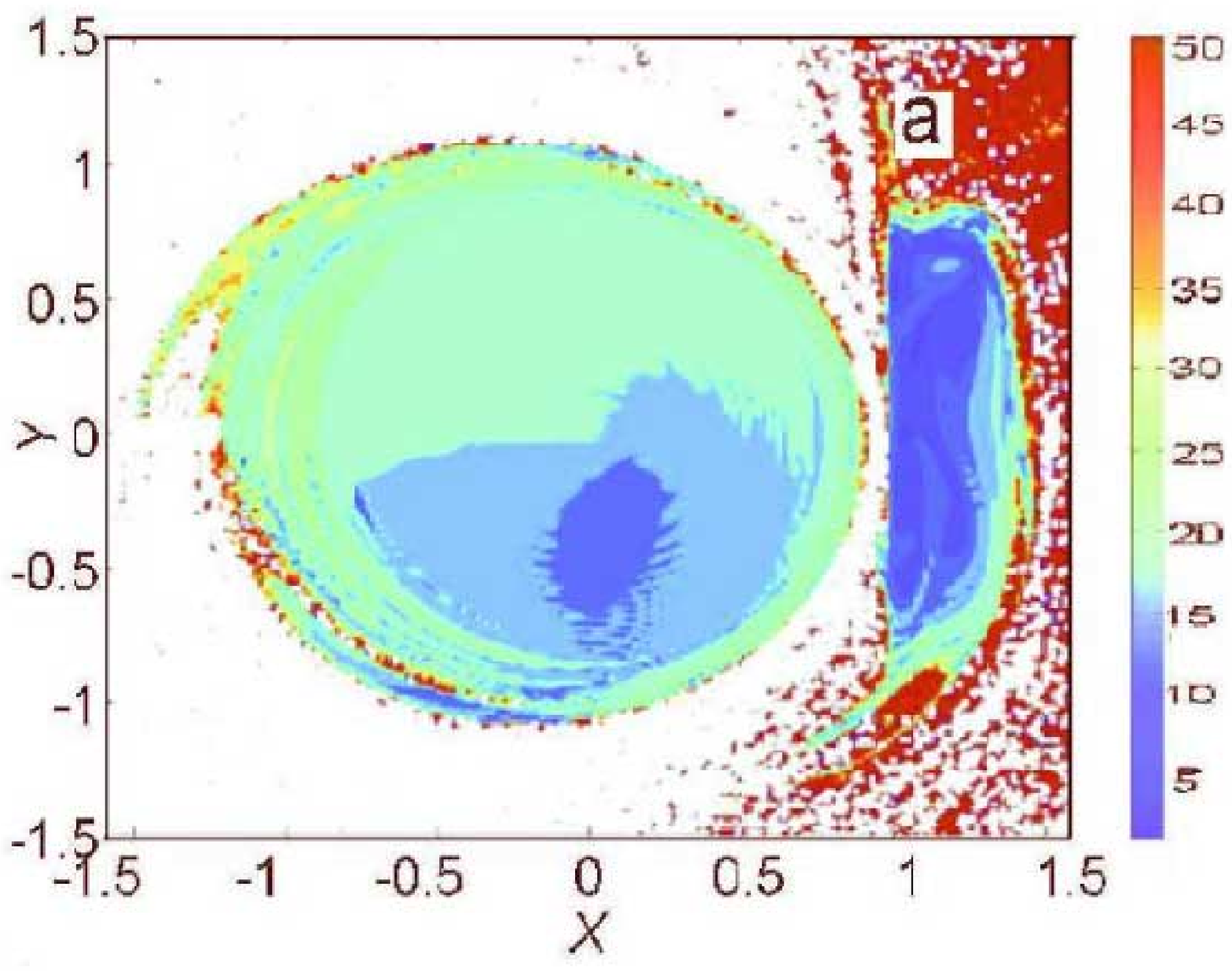}
 \includegraphics[width=5.cm,angle=-90]{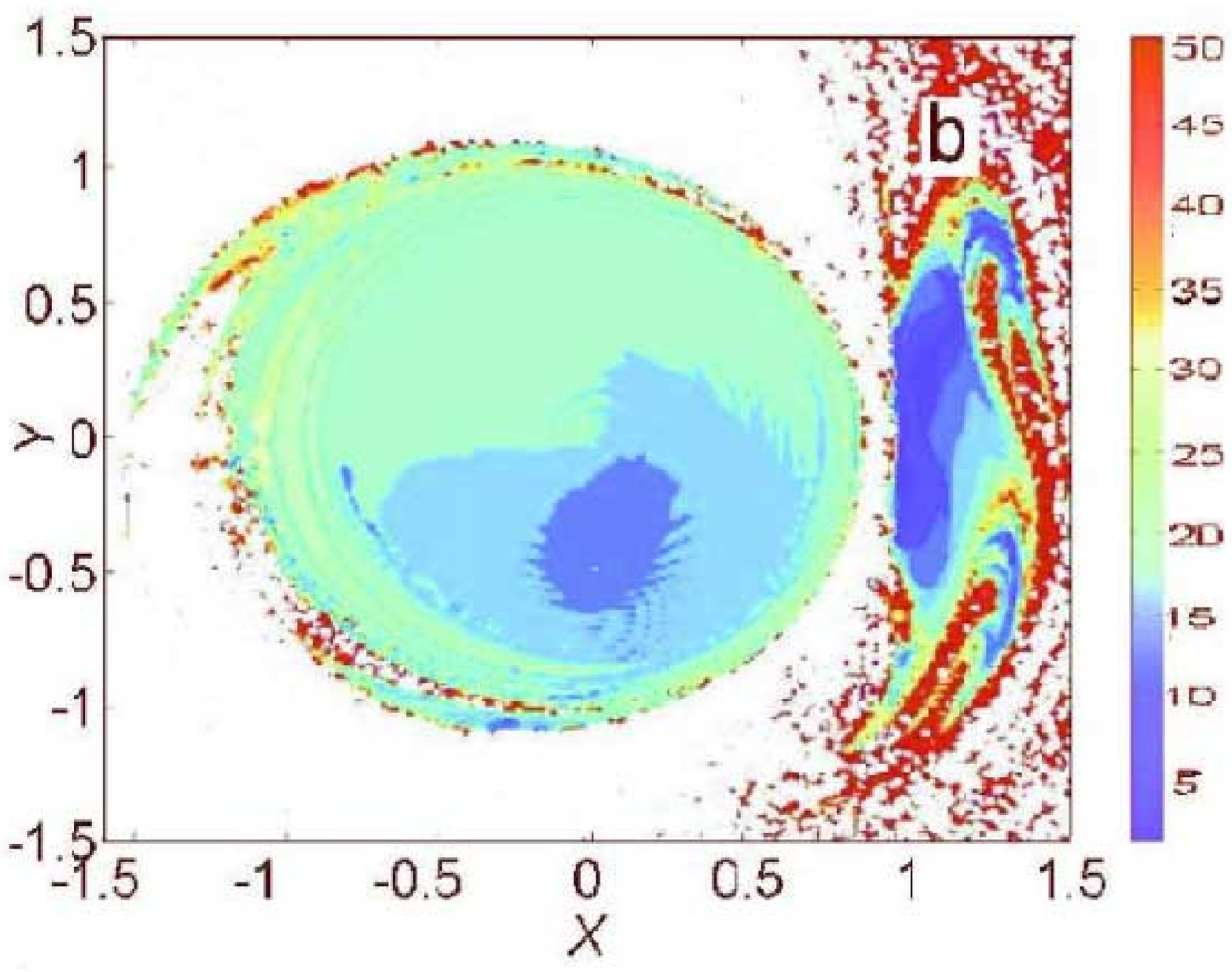}
 \includegraphics[width=5.cm,angle=-90]{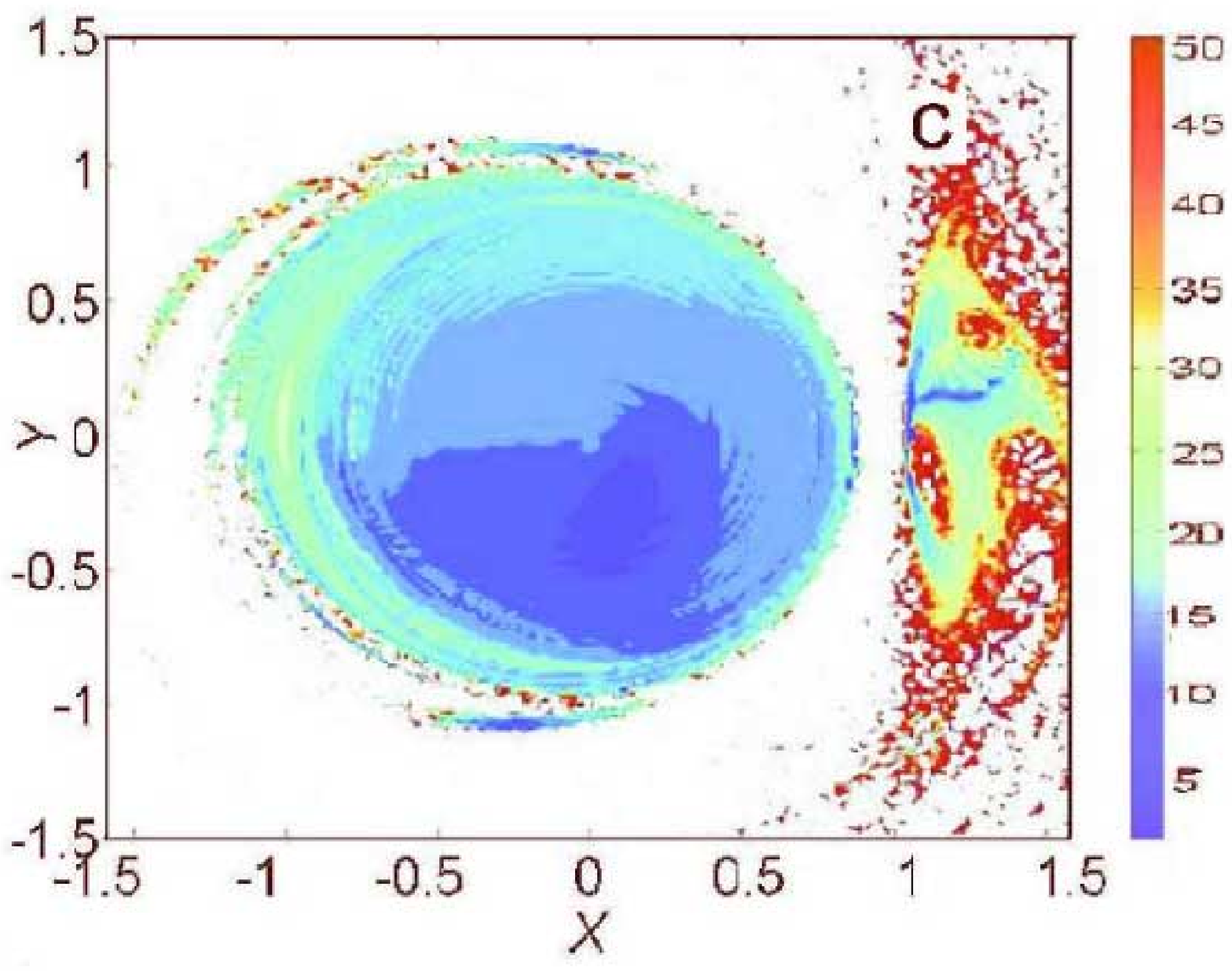}
 \includegraphics[width=5.cm,angle=-90]{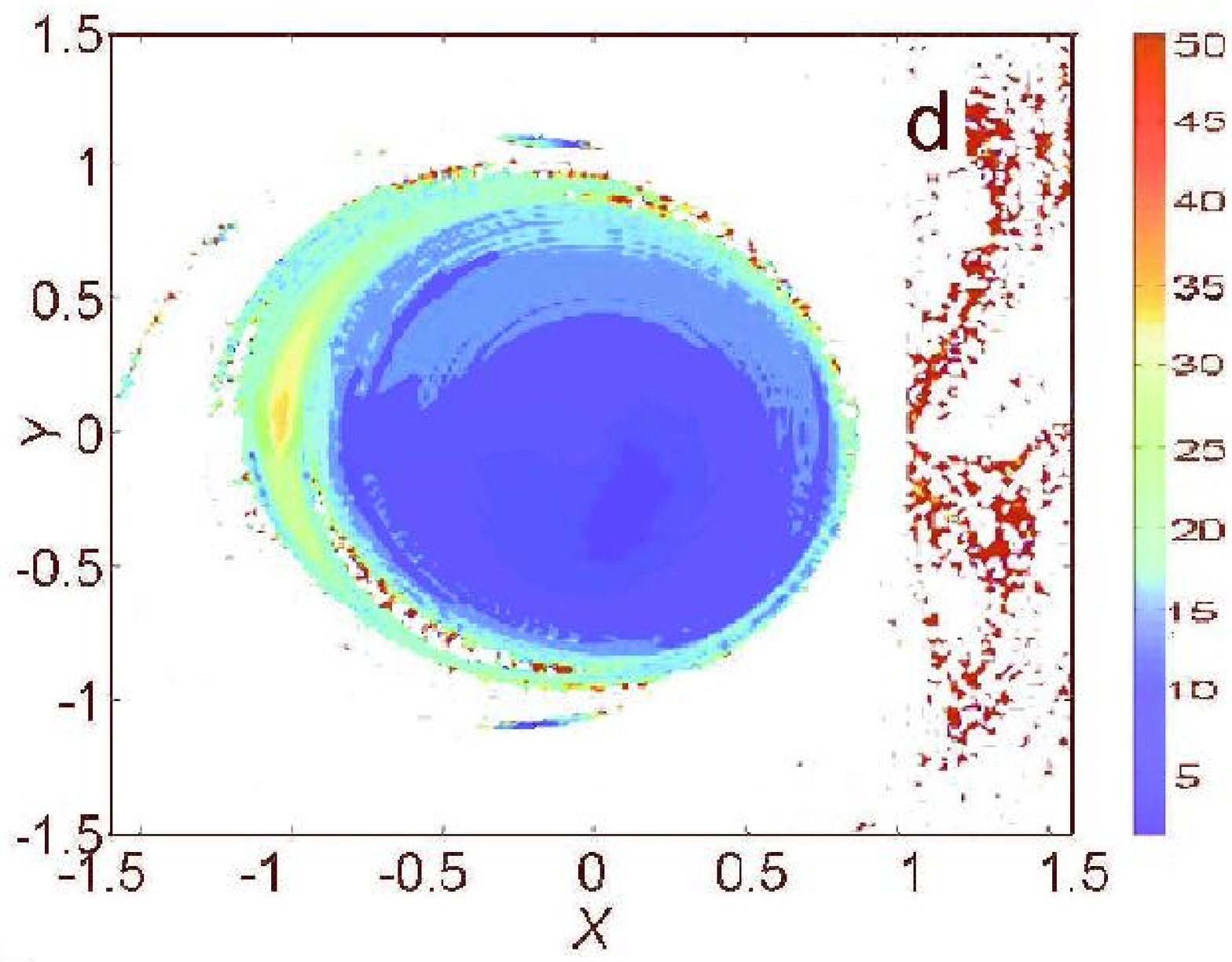}
 \end{center}
 \caption {\label{fig:fig3} (Color online) FLI contour plots for hydrogen in a LP microwave field.
 Initial energy is equal to the maximum of the ZVS
 $K_{max}=-1.3807,B=0$.
 The scaled amplitudes of the electric field $F$ are: a) 0.117, b) 0.13, c) 0.16, and d) 0.2.}
\end{figure}

In this section the FLI stability results are presented for the LP
field case. The magnetic field is zero. For each initial condition
from the subspace defined in the previous section Eq.~$(10)$ is
integrated for a time $t=100$ (the time unit is one picosecond).
The value of $\phi(t)$ in Eq.~$(11)$ was evaluated at each
moment of time. The maximum value attained over the integration
interval $[0, 100]$ is mapped onto the FLI plot. In
Fig.~\ref{fig:fig3} FLI contour plots are shown for several
distinct amplitudes of the scaled LP electric field. The color
code is assigned according to the maximum value of the FLI
evaluated for each trajectory. The dark (blue in color version)
color corresponds to the low values of the FLI and hence regular
behavior; meanwhile the light (yellow and red) color indicates
high values of the FLI and hence chaotic behavior. All the
trajectories that ionize quickly are discarded and not marked on
the plot (white regions). In addition, strongly chaotic
trajectories with the value of the FLI greater than the critical
value $\phi_c>50$ were discarded. In most cases strongly chaotic
trajectories with $\phi(100)>50$ escape to infinity and ionize
over the finite--time interval.

In the FLI contour plots presented in Fig.~\ref{fig:fig3} two
islands of stable motions can be clearly distinguished. In fact, a
similar structure of the FLI plots exists for amplitudes of the
electric field in the interval $[0.117, 0.2]$. The central and
right islands are located at the center and at the maximum of the
ZVS respectively (see Fig.~\ref{fig:fig1}). The structure of the
central island does not change with increasing amplitude of the
field. This island is constituted by the bounded states that
remain stable in a given range of the field amplitude. A layer of
chaotic motions is located around the edges of the main stable
islands. By comparing the FLI plots on Fig.~\ref{fig:fig3} with
the zero--velocity contour plot in Fig.~\ref{fig:fig1} we observe
that the stable island on the right hand side is located exactly in the
vicinity of the Lagrange maximum of the ZVS. The area and the
structure of the island change with increasing amplitude. The island
remains almost the same for the field amplitudes $F=0.117$ and
$F=0.13$. It shrinks at a field amplitude $F=0.16$ and
completely disappears at $F=0.2$. These changes are the
consequence of the break-up of invariant tori within the
island. With increasing amplitude of the field they open paths for
chaotic orbits to escape from the vicinity of the Lagrange maximum
and to ionize.

\begin{figure}
 \begin{center}
 \includegraphics[width=5.cm,angle=-90]{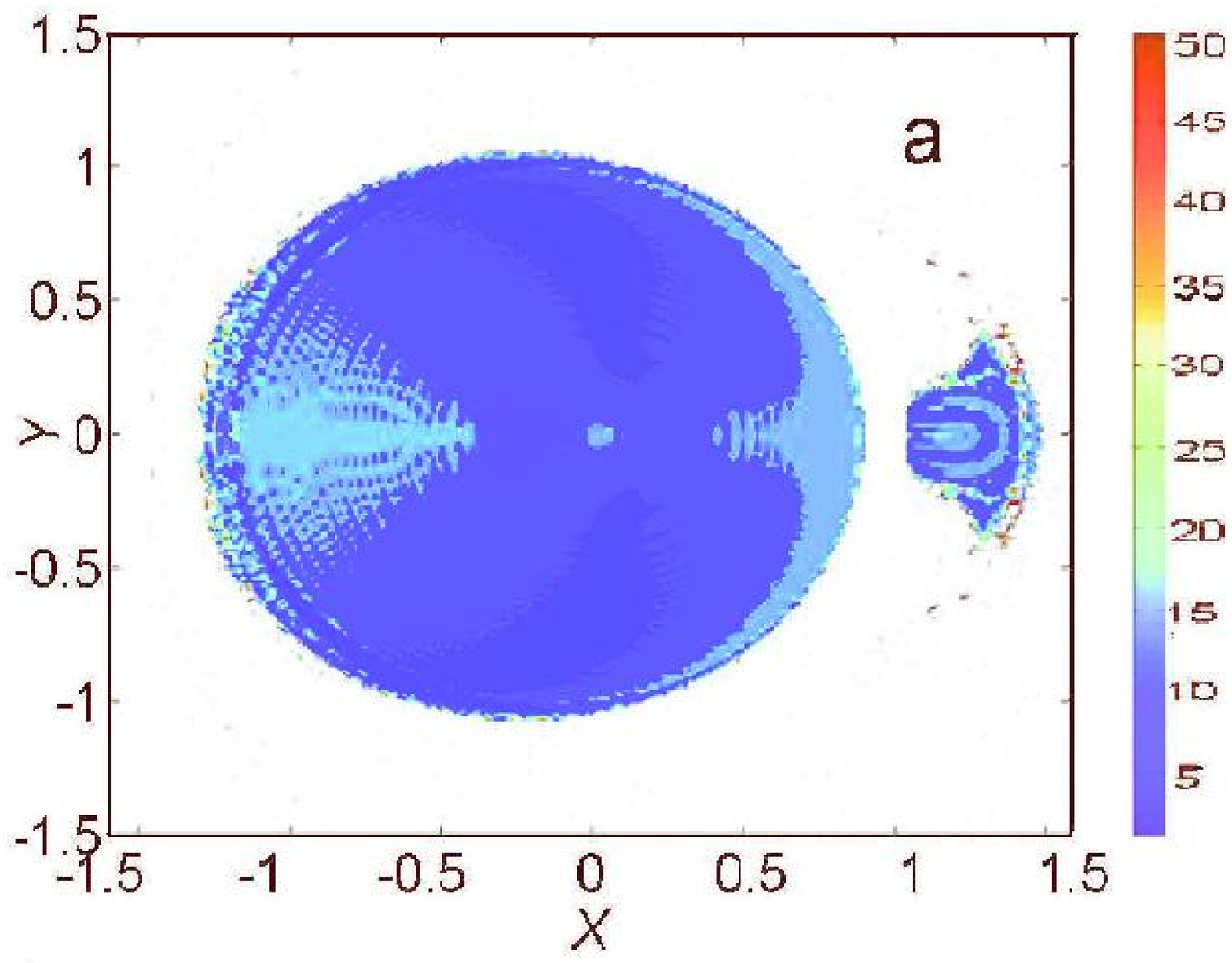}
 \includegraphics[width=5.cm,angle=-90]{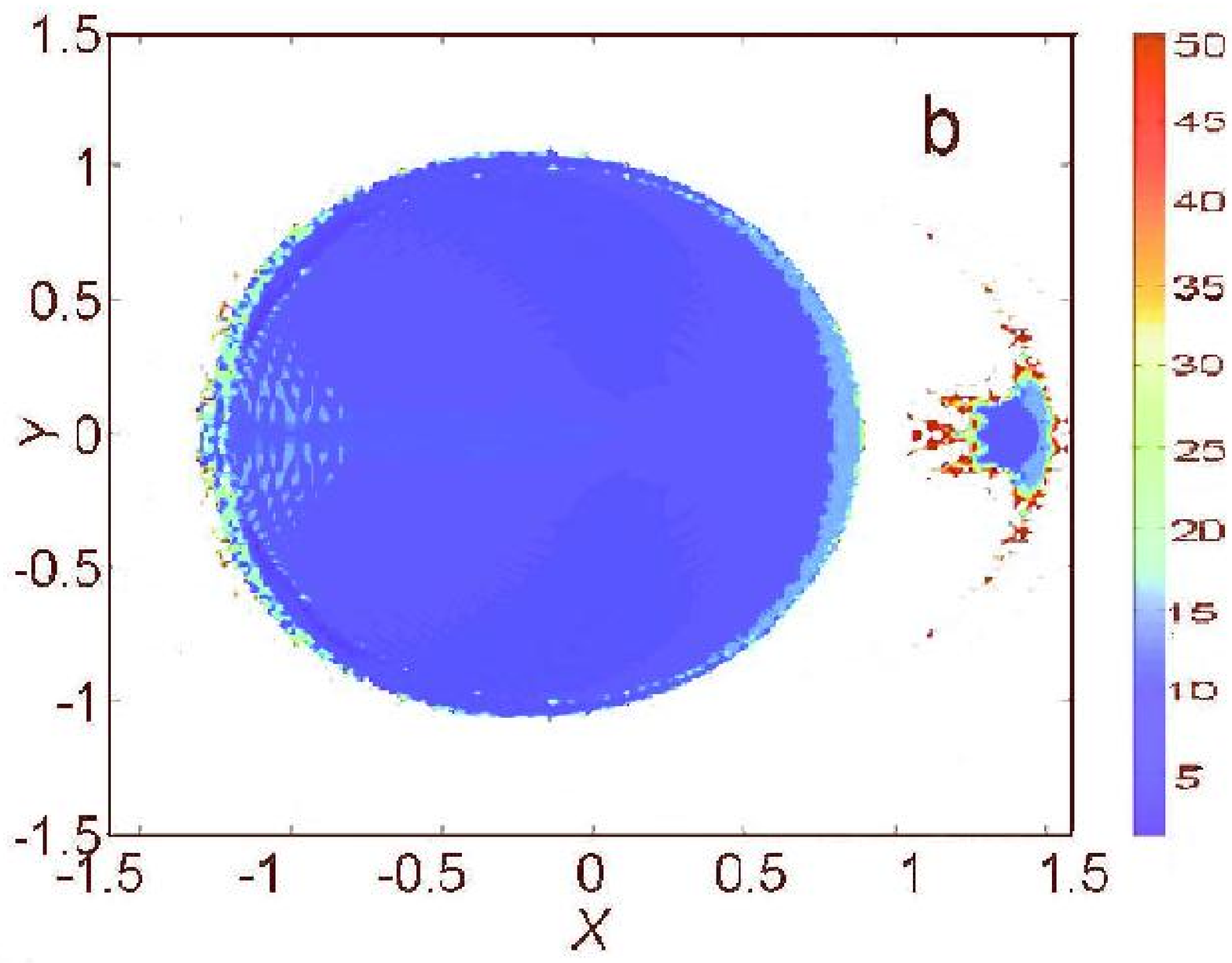}
 \includegraphics[width=5.cm,angle=-90]{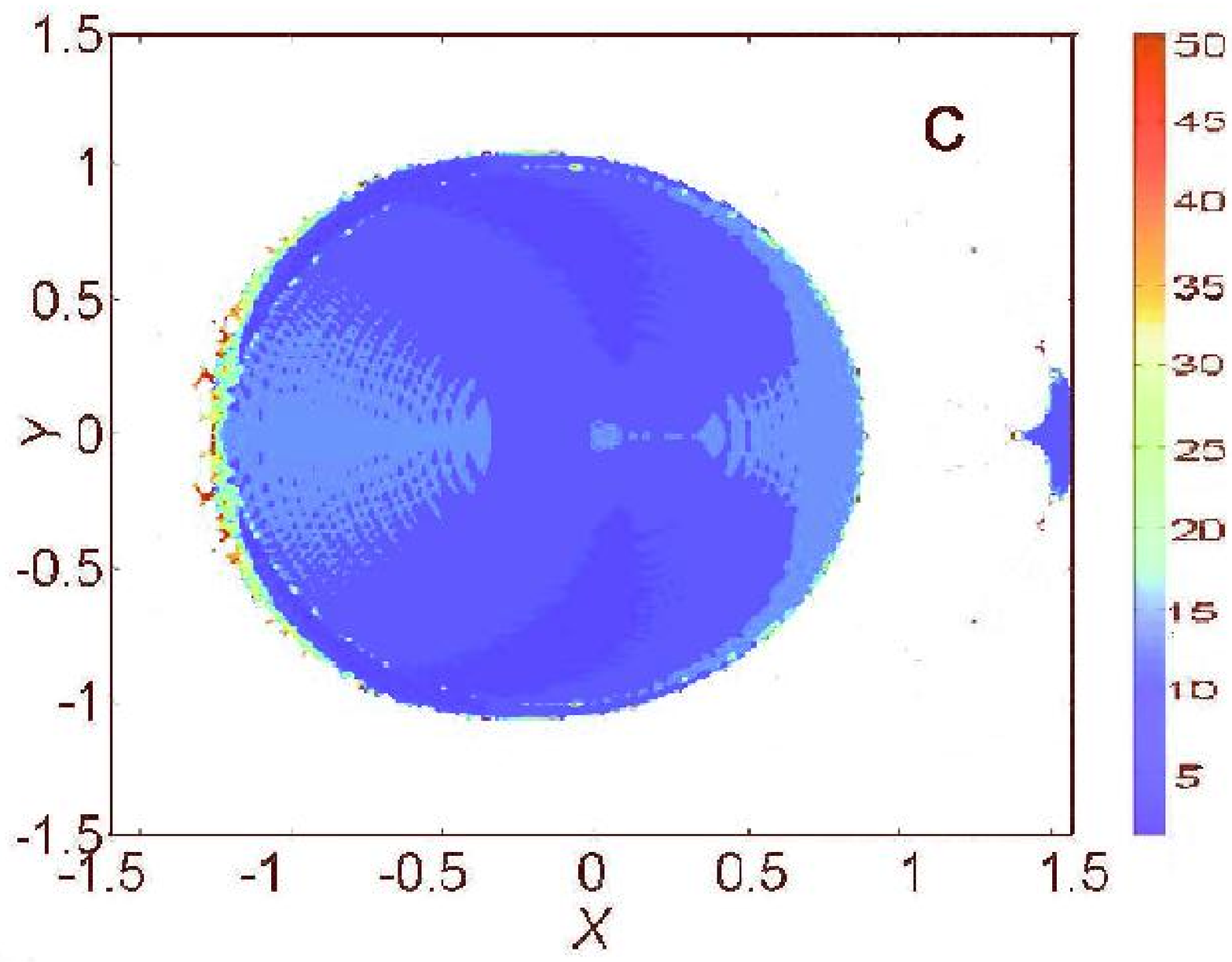}
 \includegraphics[width=5.cm,angle=-90]{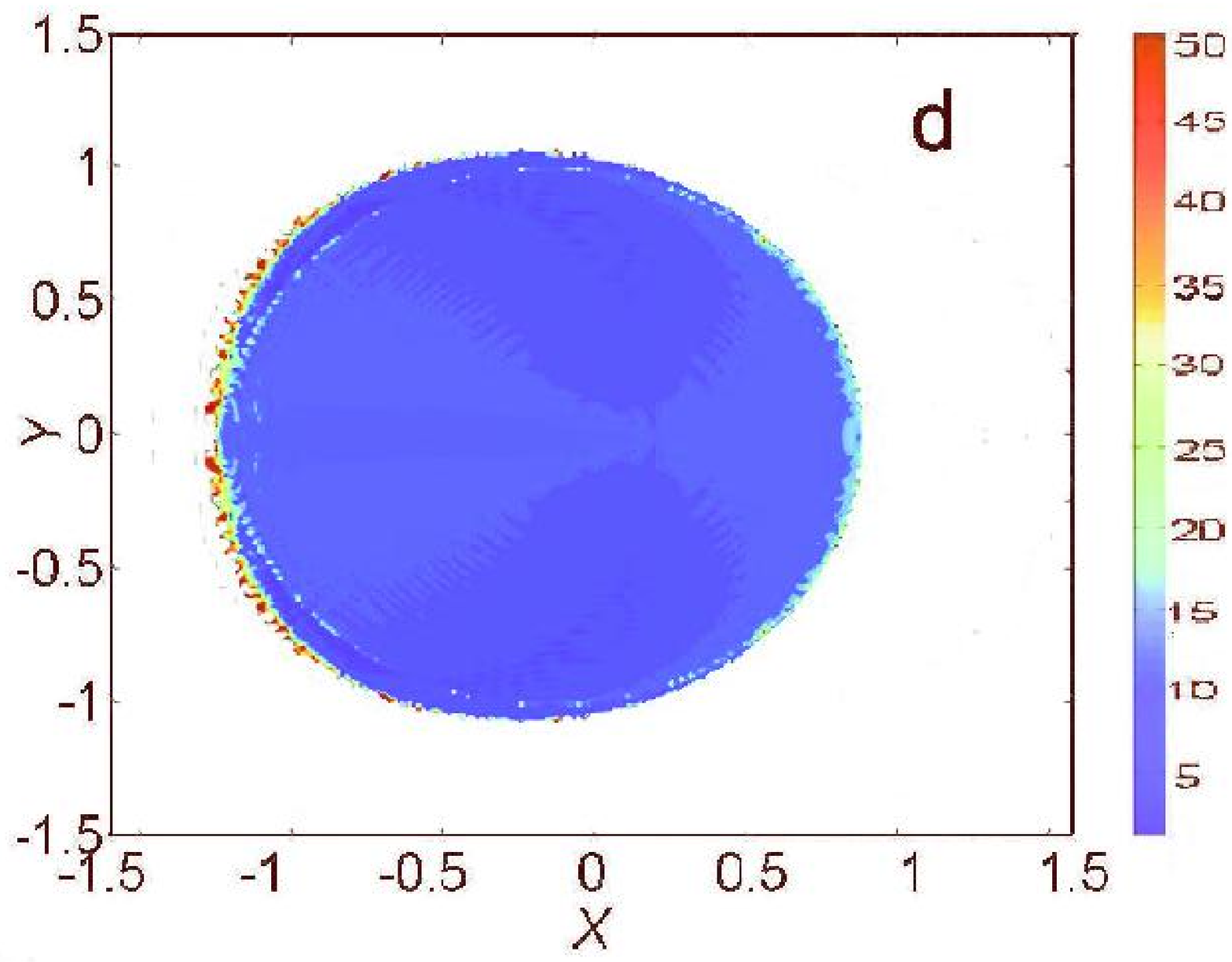}
 \end{center}
 \caption {\label{fig:fig4} (Color online) FLI contour plots for hydrogen in  a CP microwave
 field. Initial energy is equal to the maximum of the ZVS
 $K_{max}=-1.3807,B=0$.
 The color scale is assigned according to the values of the FLI.
 The scaled amplitudes of the electric field
 $F$ are: a) 0.117, b) 0.13, c) 0.14, and d) 0.15.}
\end{figure}
\begin{figure}
 \begin{center}
 \includegraphics[width=5.cm,angle=-90]{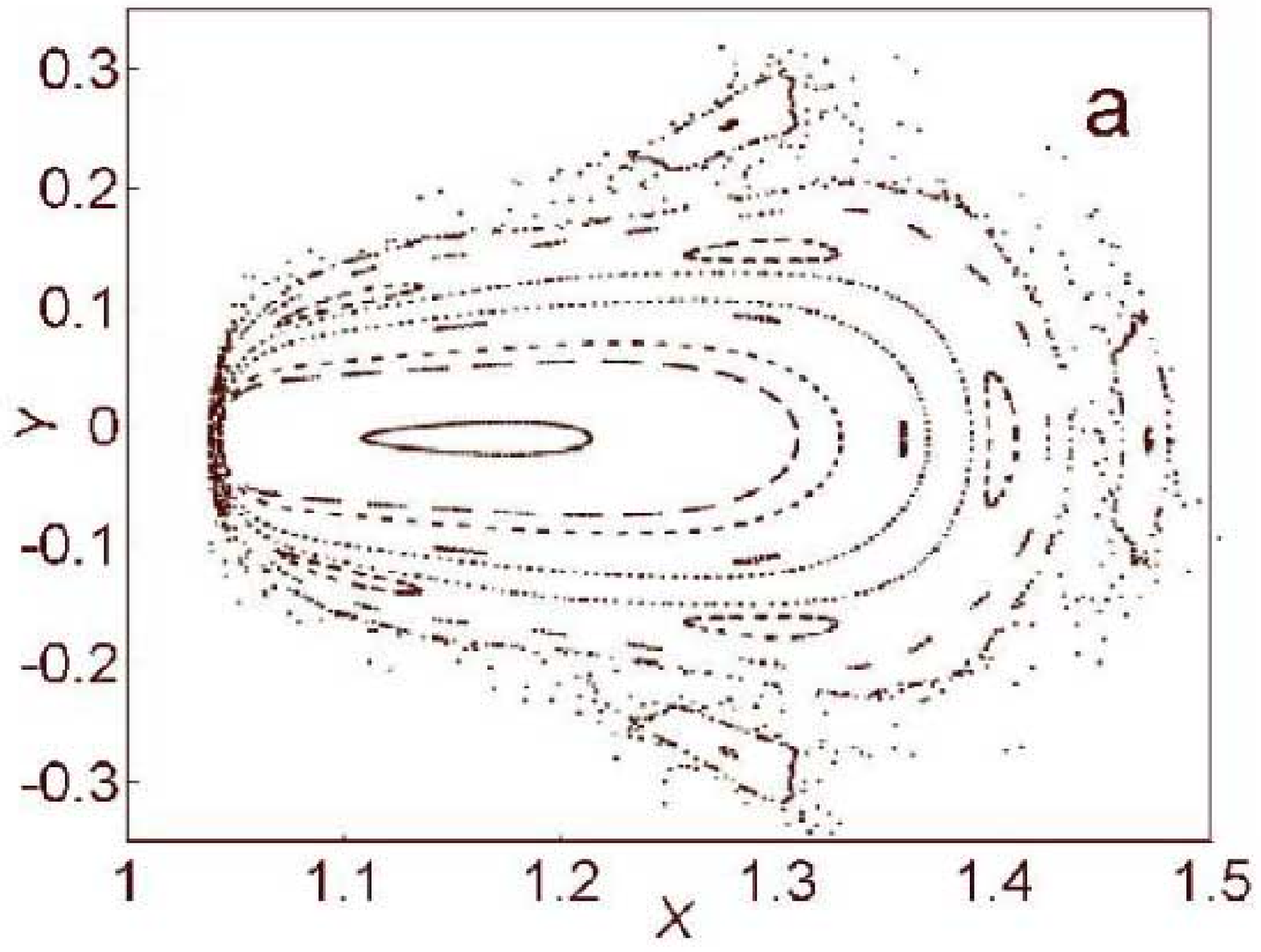}
 \includegraphics[width=5.cm,angle=-90]{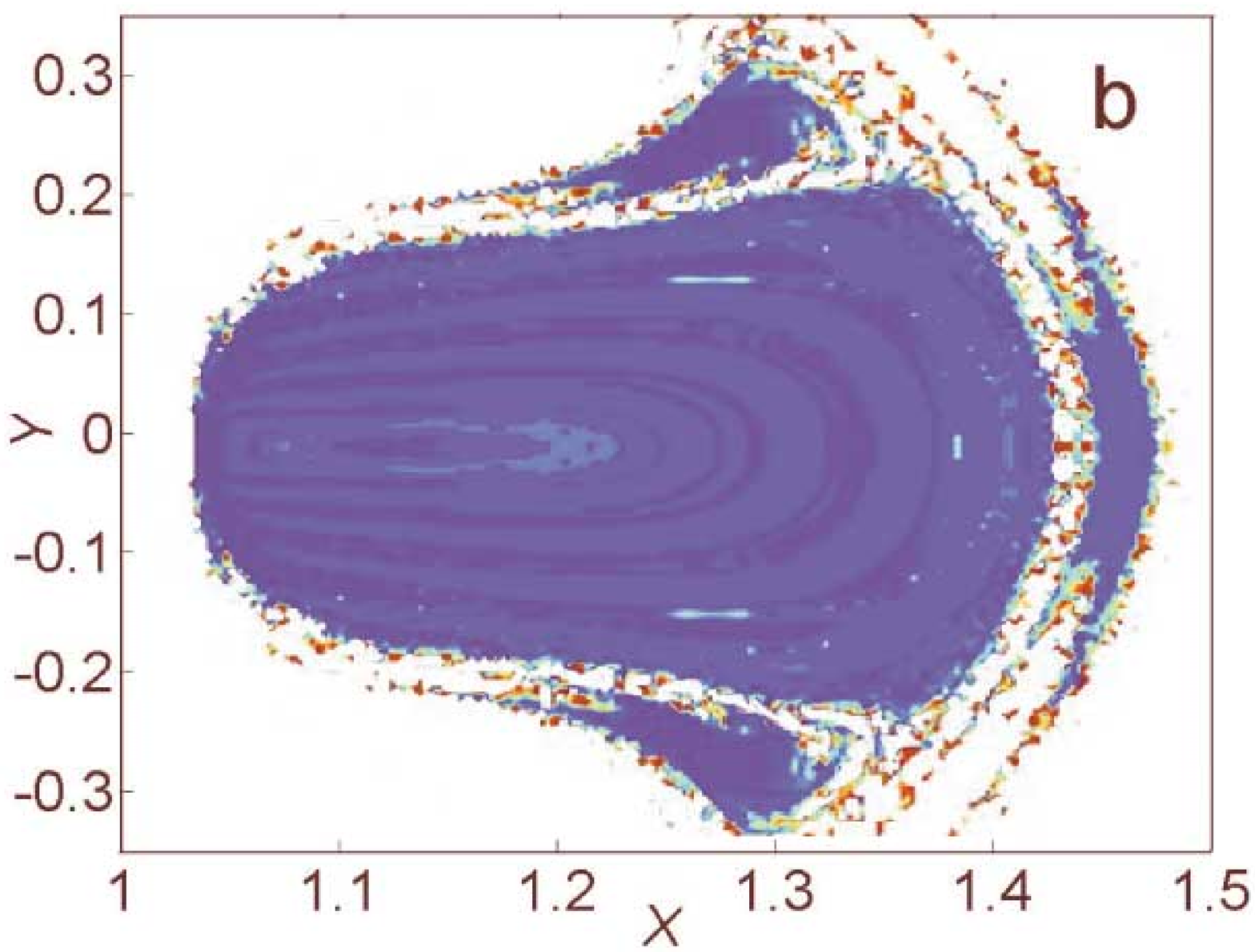}
 \end{center}
 \caption {\label{fig:fig5} (Color online) The Poincar\'{e} surface of section and the
 magnification of the small island from Fig.~\ref{fig:fig4}$(a)$
 are shown on panels $(a)$ and $(b)$ respectively.}
\end{figure}

In Fig.~\ref{fig:fig4} the FLI plots are shown for the CP problem.
First, one observes that in all the four plots the size of the
stability island around the Lagrange maximum is much smaller than
the size of similar structures for the LP problem. Second, the
size and structure of the central island look exactly the same as
for the LP case. For the CP problem, the bounded states inside the
central island remain stable for all the field amplitudes from the
interval $[0.117, 0.2]$. The field amplitude for which all the
trajectories launched from the right hand side island ionize ($F=0.15$) is
lower than the corresponding amplitude ($F=0.2$) for the LP case.
As was mentioned before, for the CP field the system is
time--independent in the rotating frame. The structures of the
FLI contour plot coincide with the ones of the Poincar\'{e}
section. As an example, in Fig.~\ref{fig:fig5} the Trojan
bifurcation is shown on the Poincar\'{e} section ($(a)$ panel) and
on the FLI contour plot ($(b)$ panel). Figure~\ref{fig:fig5}$(b)$
is the magnification of the small island of
Fig.~\ref{fig:fig4}$(a)$.
 A one--to--one correspondence of
both plots is evident. On a close inspection, the symmetry about
the $y$ axis is observed. The resonant zones on the Poincar\'{e}
section correspond to the dark (blue in color version) islands on
the FLI plot. The chaotic zones are marked around the edges of the
islands.

\begin{figure}
 \begin{center}
 \includegraphics[width=5.cm]{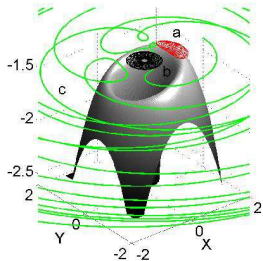}
 \end{center}
 \caption {\label{fig:fig6} (Color online) Two stable orbits $(a)$, $(b)$, and one ionizing
 orbit $(c)$ versus the zero--velocity surface. The parameters are $K=K_{max}, F=0.117, B=0$, and $\alpha=1$.}
\end{figure}

To illustrate the FLI stability results, we show the ionization
dynamics for the trajectories with the initial conditions inside
two stable regions identified on the FLI plots and one trajectory
with initial conditions inside the chaotic region. In
Fig.~\ref{fig:fig6} the configurations of three trajectories are
shown as well as ZVS surface in the rotating frame. Initial
conditions are chosen from different regions in
Fig.~\ref{fig:fig4}: $(a)$ from the big stable island, $(b)$ from
the small stable island, and $(c)$ from the chaotic region.
Trajectories $(a)$ and $(b)$ correspond to the low eccentricity
bounded states and trajectory $(c)$ is chaotic and ionizes after
several close encounters with the nucleus.

\subsubsection{FLI analysis for EP field}
\label{sec:4.1.3}

\begin{figure}
 \begin{center}
 \includegraphics[width=5.cm,angle=-90]{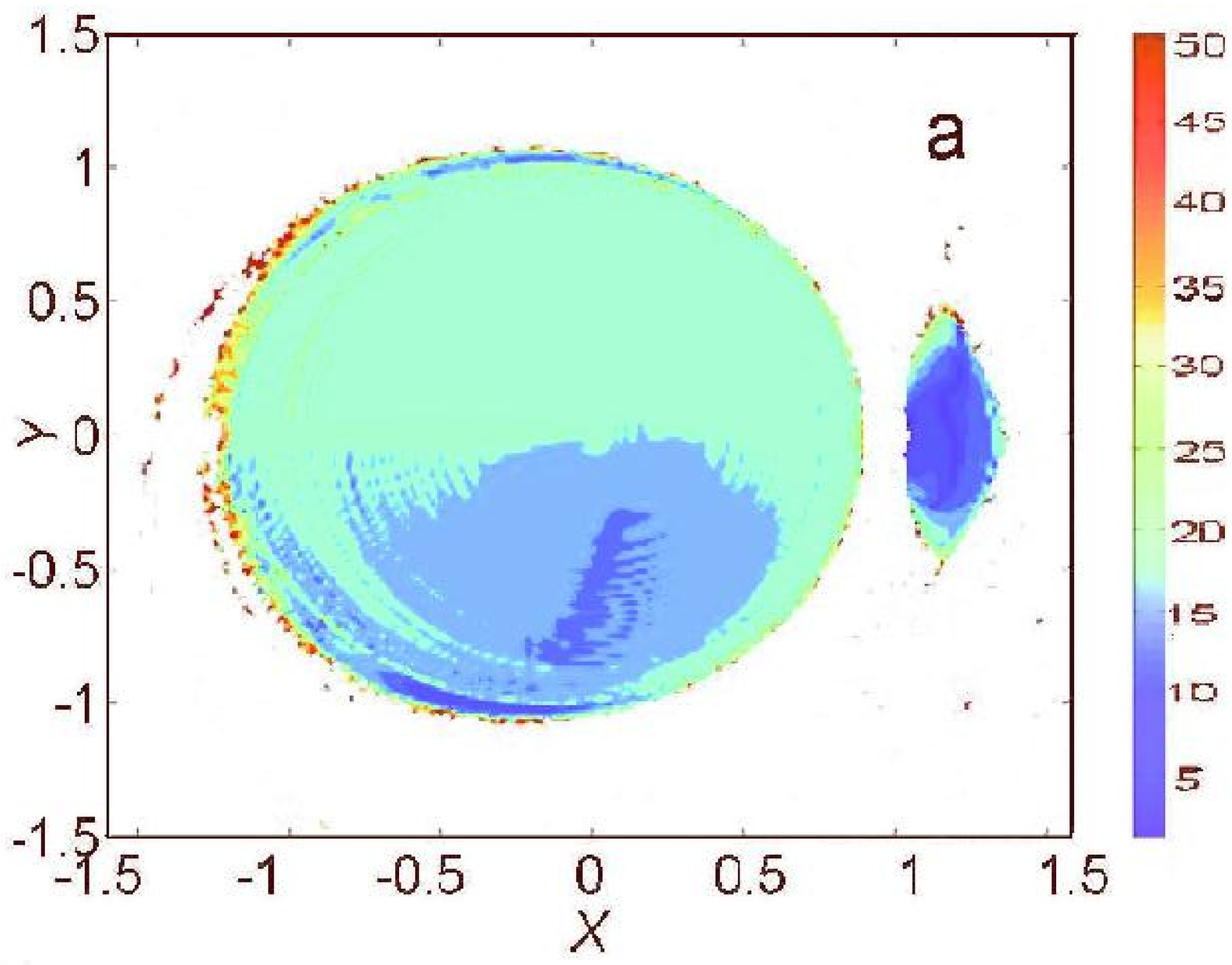}
 \includegraphics[width=5.cm,angle=-90]{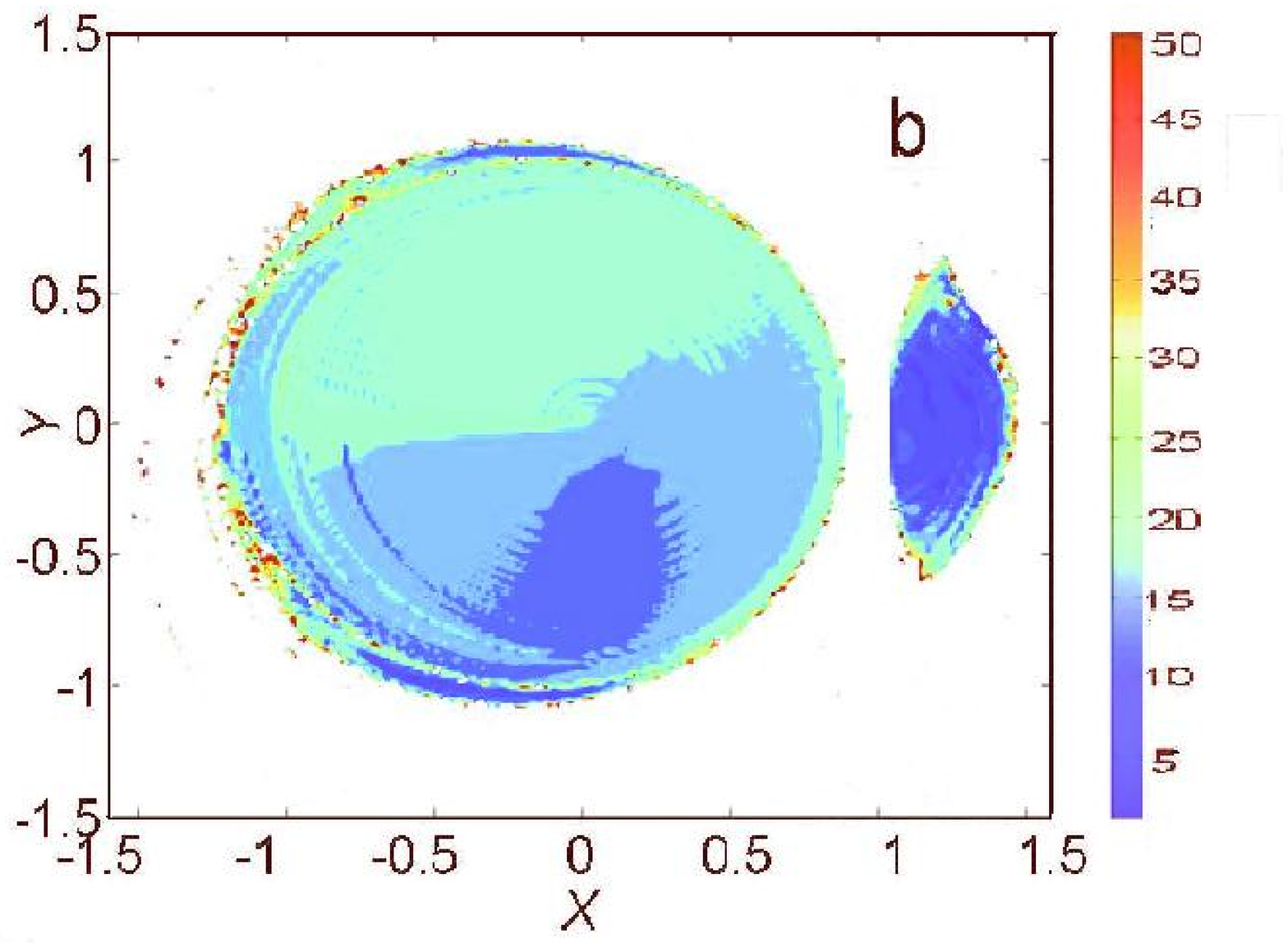}
 \includegraphics[width=5.cm,angle=-90]{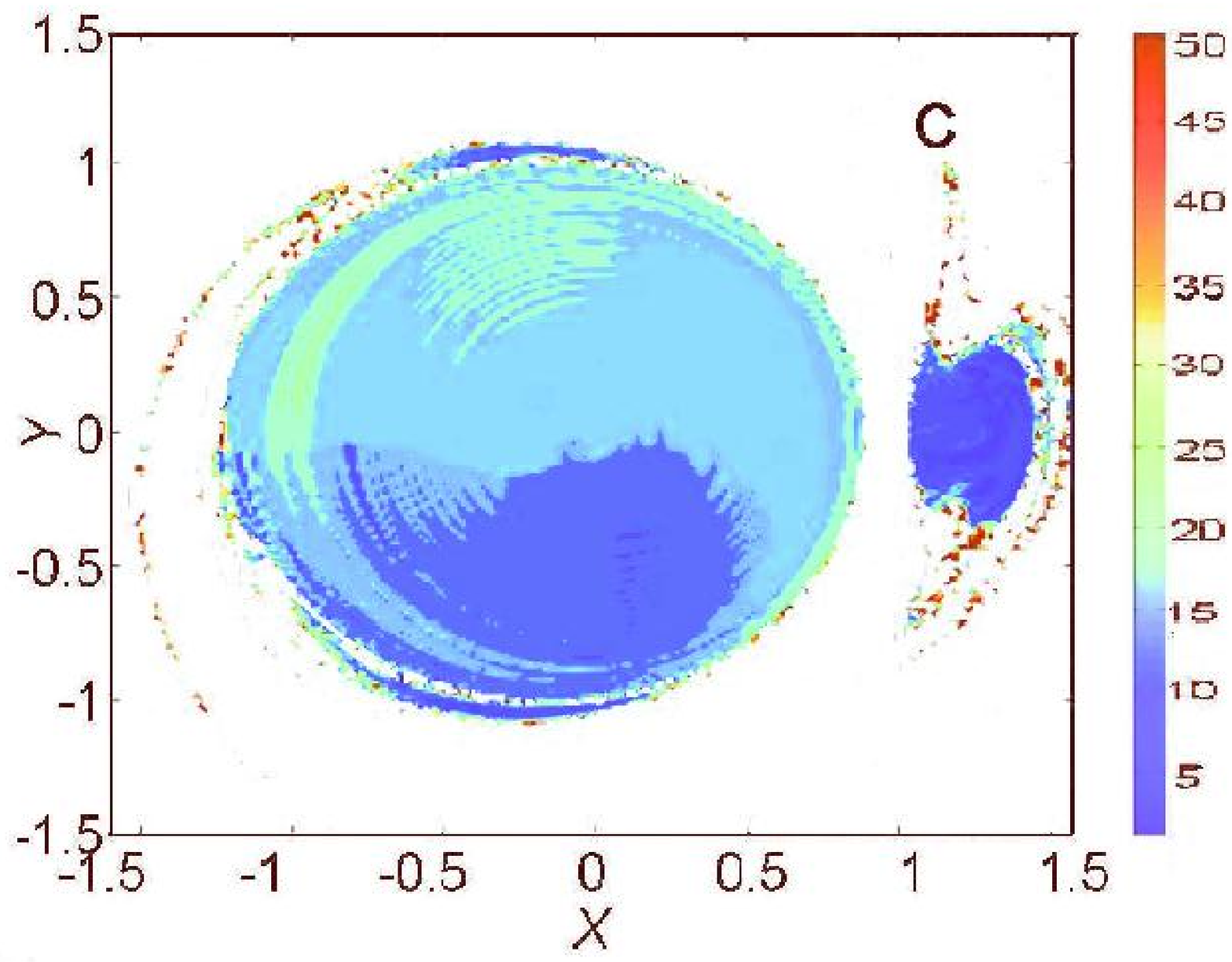}
 \includegraphics[width=5.cm,angle=-90]{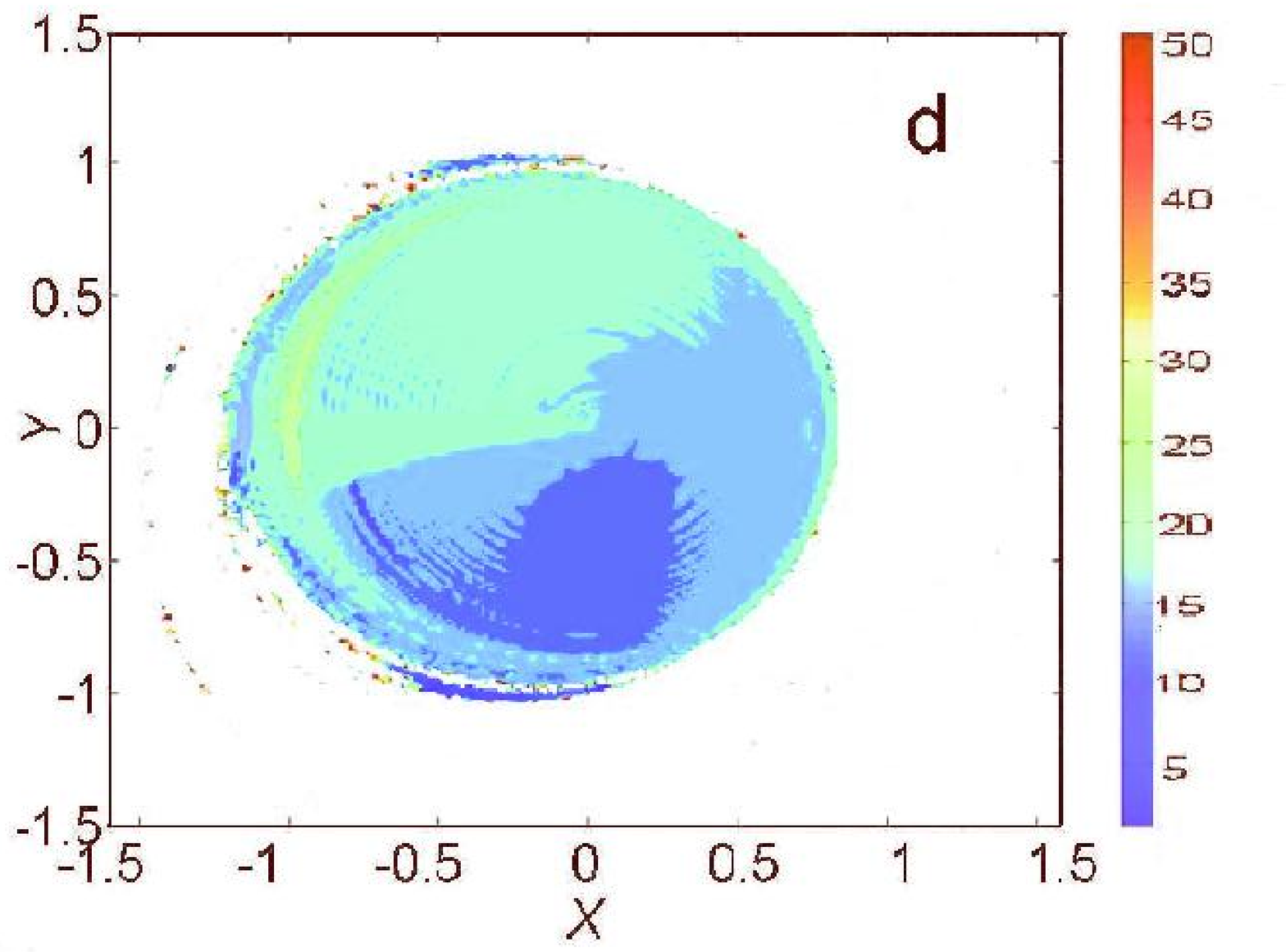}
 \end{center}
 \caption {\label{fig:fig7} (Color online) FLI stability plots for hydrogen in EP microwave field.
 $K_{0}=K_{max},\alpha=0.6,B=0$.
 The scaled amplitudes of the electric field
 $F$ are: a) 0.117, b) 0.13, c) 0.15, and d) 0.17.}
\end{figure}

The FLI plots in Fig.~\ref{fig:fig7} illustrate the structures for
the hydrogen atom in EP field for the polarization degree
$\alpha=0.6$. The scaled amplitude of the electric field is taken
within the interval $[0.117,0.17]$. In general, the structure of
the FLI plots for the EP field appears to be similar to the
stability plots for the CP field case. Two islands of stability
are apparent: a big island located at the center of ZVS and a
small island located at the maximum of ZVS (see
Fig.~\ref{fig:fig1}). The ellipticity of the field destroys the
symmetry that is present for the CP field case. It is evident that
the right hand side island in Fig.~\ref{fig:fig7}~$(c)$ is not
symmetric around the $y$ axis as opposed to the ones around the
maximum in Fig.~\ref{fig:fig4}. Lightly-colored regions can be
seen on the edges of the two islands. These are initial conditions
corresponding to chaotic trajectories with high values of the FLI.
Trajectories with the values of the FLI higher than the critical
$\phi_c=50$ are not represented. It can be observed that the large
area surrounding the stable islands corresponds to rapidly
ionizing chaotic motions. Similarly to the structure of the
central island observed in Figs.~\ref{fig:fig3} and
\ref{fig:fig4}, the equivalent structure for the EP field remains
essentially unchanged with increasing strength of the field. At
the same time noticeable changes are seen in the size and
structure of the small island. The size of the small island varies
substantially with the increase of the field amplitude. The growth
of the island at certain amplitudes is accounted for the
stabilization of some resonant motions within the island. For
example, at $F=0.13$ the size of the island is noticeably larger
than its size at $F=0.117$ in Fig.~\ref{fig:fig7}~$(a)$. At
$F=0.17$ the small island disappears, i.e.\ most of the invariant
tori within the island that existed at lower amplitudes of the
field have been broken and almost all of the trajectories in the
vicinity of the maximum become chaotic and ionize.
The FLI plots shows that the dynamics in the EP case is more
regular than the CP case. For example, the size of the stable zone
around the maximum point in Fig.~ \ref{fig:fig7} appears to be
larger than the size of the stable zone in Fig.~\ref{fig:fig4}. On
the other hand the size of the right hand side island in Fig.~\ref{fig:fig7}
for the EP field is much smaller than the size of the
corresponding island in Fig.~\ref{fig:fig3} for the LP field. From
our detailed examination of the FLI results for the initial energy
equal to the maximum of ZVS we conclude that the phase space
dynamics is more regular for the LP field case than for the EP and
CP field cases. Moreover, the size of the stable region around the
Lagrange maximum point is observed to vary non--monotonically with
the increase of the field amplitude for the EP field case. These
changes are ascribed to various nonlinear resonant effects that
were studied classically in Ref.~\cite{rich97} and quantum
mechanically using Floquet state approach in Ref.~\cite{oks99}.

\subsection{Study of ionization probability curves}
\label{sec:4.2}

It has been pointed out in Ref.~\cite{farrelly95} that the
apparent ionization threshold for Rydberg atoms in the CP
microwave field must be determined by the fraction of orbits that
undergo first transition to chaos. Such orbits are
located outside of the ZVS and they represent the atomic states
that can be easily populated in the experiment \cite{beller97}. In
connection with these results we found the set of bounded orbits
in the vicinity of the Lagrange maximum of the ZVS. These are the
orbits that undergo the transition from a regular to chaotic
behavior within the range of electric field amplitude $F\in[0.12,
0.18]$. The FLI plots in Figs.~\ref{fig:fig3},~\ref{fig:fig4} and
\ref{fig:fig7}, which are computed for the linear ($\alpha=0$),
circular ($\alpha=1$), and intermediate ($\alpha=0.6$)
polarizations, illustrate the changes of the stability of these
orbits. While the dynamics in the central island remain unaffected by
the increase of the field, the orbits located near the Lagrange
maximum become chaotic and ionize with the increase of the field
amplitude. For this reason, we determine an apparent ionization
threshold by measuring the fraction of chaotic orbits from the
phase space volume enclosing the Lagrange maximum. The
calculations of the ionization probabilities are carried out for
the polarizations of the field from the circular to the linear
limit ($0<\alpha<1$). The behavior of the ionization probabilities versus
scaled electric field amplitude $F\in[0.12,0.18]$ is estimated by
means of the FLI analysis.

 By monitoring the evolution of the FLI along each
integrated trajectory we count the number of chaotic trajectories
with the FLI value equal or above the critical value $\phi_c=50$.
All chaotic trajectories are considered as possible candidates
for the classical ionization. Indeed, the absence of invariant
tori outside of the two stable structures distinguished by the FLI
analysis shows that all chaotic trajectories escape from the
vicinity of the Lagrange maximum to a zone associated with the
high classical action values. The FLI plots in
Figs.~\ref{fig:fig3},\ref{fig:fig4} and \ref{fig:fig7} show that
the phase space surrounding the island at the Lagrange
maximum is chaotic. There are no stable structures (except the
big stable island at the center) that can prevent the escape of
chaotic trajectories and their subsequent ionization.

\begin{figure}
 \begin{center}
 \includegraphics[width=6cm]{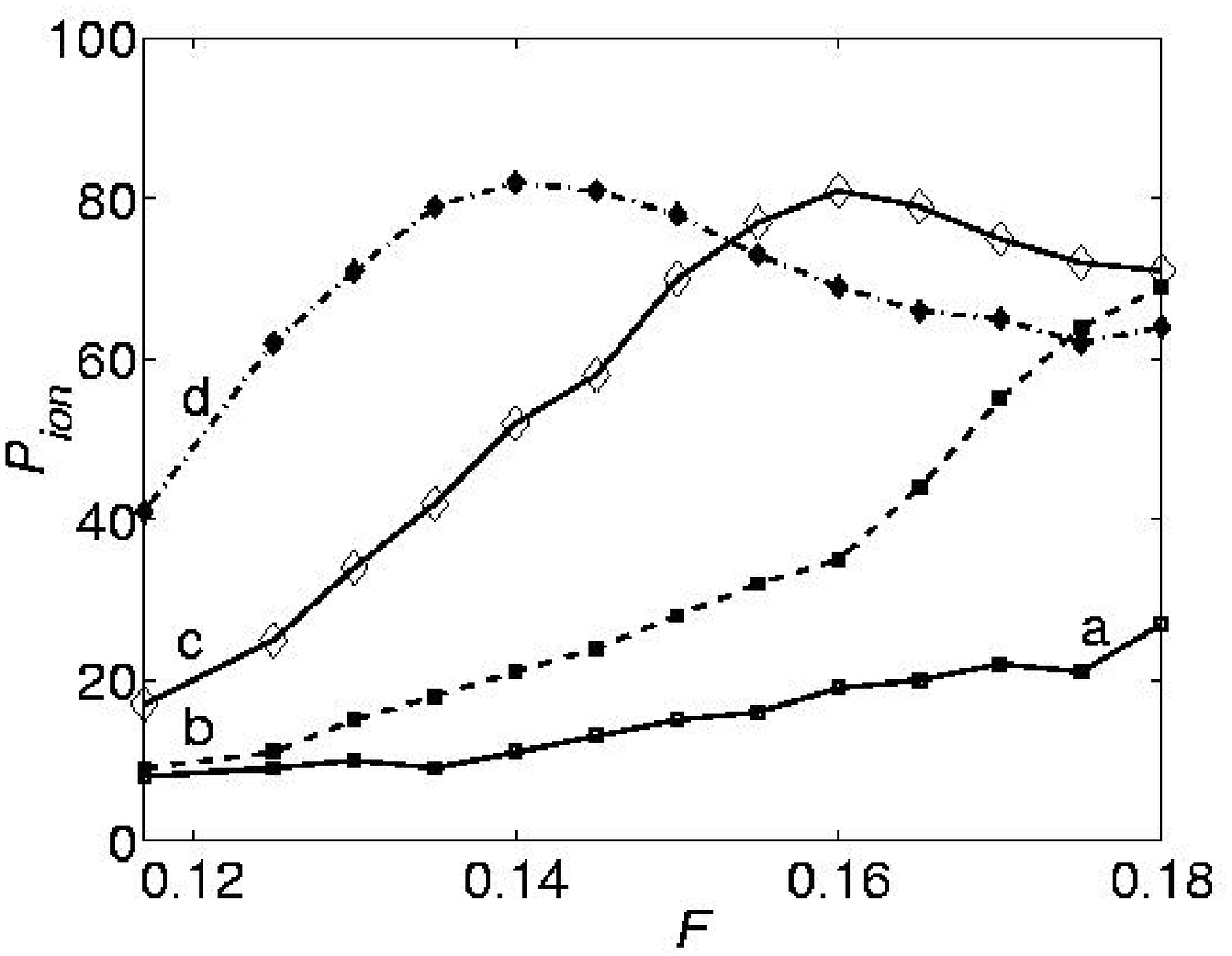}
 \includegraphics[width=6cm]{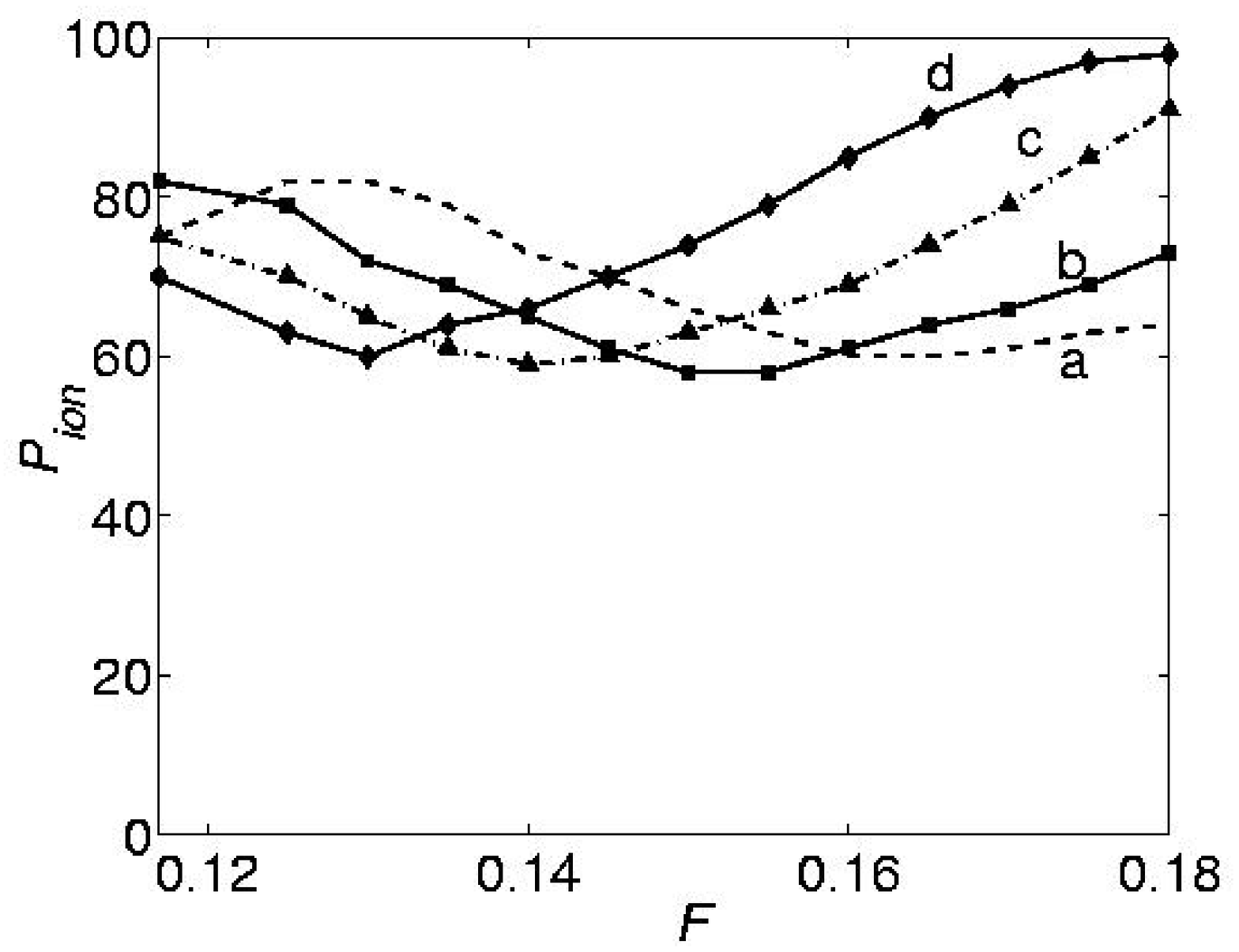}
 \includegraphics[width=6cm]{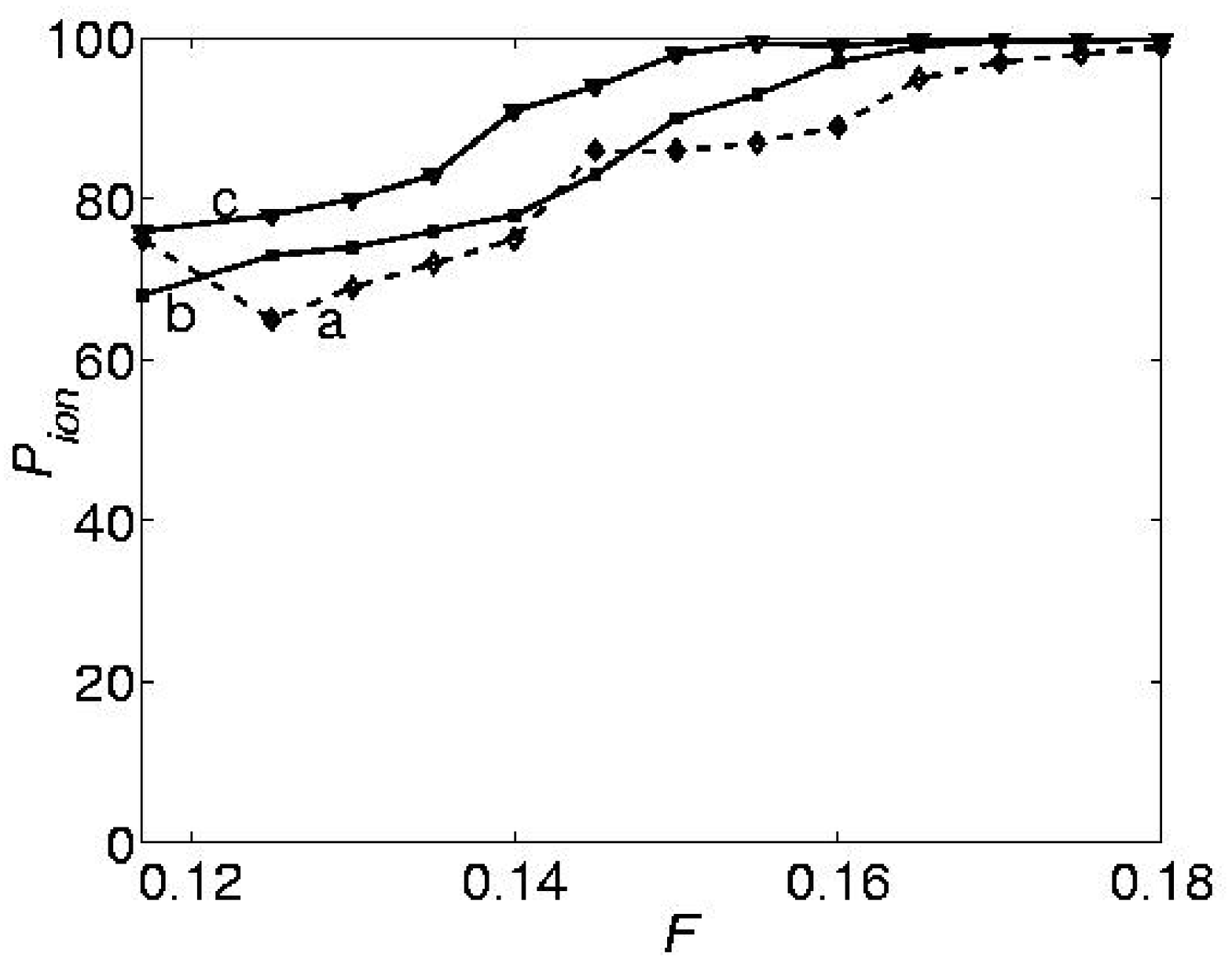}
 \caption {\label{fig:fig8}
Ionization probability $P_{ion}$ computed for the ensemble of
trajectories of hydrogen in EP microwave electric field. The
parameters are $B=0, K(0)=K_{max}=-1.3807$. Top: ionization
probabilities for the polarizations close to LP: a) 0, b) 0.1, c)
0.2, and d) 0.3. Center: Ionization probabilities for the
intermediate polarizations: a) 0.4, b) 0.5, c) 0.6, and d) 0.7.
 Bottom: Ionization probabilities for the polarizations
close to CP: a) 0.8, b) 0.9, and c) 1.}
\end{center}
\end{figure}

We define the percentage ionization probability from the ratio of the number of
chaotic trajectories $N_{chaot}$ to the total number of
trajectories $N_{total}$ from the volume of phase space
surrounding the maximum point  as follows:

%$P_{ion}=N_{chaot}/N_{total}\times
%100 \%$.
\begin{equation}
P_{ion}=\frac{N_{chaot}}{N_{total}}\times 100 \%.
\end{equation}

The resulting apparent ionization probabilities versus the scaled field
amplitude are shown in Fig.~\ref{fig:fig8}. Each curve represents the
ionization probability for different field polarizations $\alpha$.
On the top panel of Fig.~\ref{fig:fig8} results are
shown for the polarizations of the field close or equal to the LP
limit. The curve $(a)$ indicates the percentage of chaotic
trajectories in the vicinity of Lagrange maximum for the LP field.
It is clearly seen that on average the percentage of chaotic
trajectories within the phase space volume around the Lagrange
maximum point is below $20\%$. The ionization probability $(a)$
increases slowly with the increase of the amplitude of the field.
It is evident, that for the LP limit the dynamics around the
Lagrange maximum remains predominantly regular for the electric
field amplitude $F$ in the interval $[0.12, 0.18]$. A qualitative
illustration of the changes in dynamics for the LP limit is given
in Fig.~\ref{fig:fig3}. From this figure one observes that
transition from regular to chaotic dynamics occurs at $F=0.2$. On
the top panel of Fig.~\ref{fig:fig8} sharp changes in the behavior
of the probabilities $(c)$ and $(d)$ are observed for the polarizations
$\alpha=0.2$ and $\alpha=0.3$ respectively. First, the probabilities
$(c)$ and $(d)$ rise rapidly from the $20\%$ and $40\%$ for
$F=0.117$ to $80\%$ for the higher field amplitudes. Secondly,
ionization probabilities $(c)$ and $(d)$ exhibit non--monotonic rise with
the increasing field amplitude. Two pronounced local maxima in the
ionization probabilities are observed for $F=0.16,~\alpha=0.2$ and for
$F=0.14, ~\alpha=0.3$. The main feature of the ionization probabilities
seen on top panel is the sharp transition from almost monotonic
behavior for $\alpha=0$ and $\alpha=0.1$ to the non--monotonic
behavior for $\alpha=0.2$ and $\alpha=0.3$. This shows that
dynamics around the Lagrange maximum strongly depends on the field
polarization.

On the central panel the ionization probabilities are shown for
intermediate polarizations. The behavior of the ionization probabilities
is non--monotonic. For instance, there is a local maximum at
$F=0.13$ and local minimum at $F=0.162$ for the curve $(a)$. The
percentage of chaotic trajectories in the vicinity of the Lagrange
maximum is equal to $80\%$. In fact, for the field amplitude
$F=0.18$ and polarization $\alpha=0.6$ the percentage of chaotic
trajectories is almost $90\%$, which means that most of the orbits
around the Lagrange maximum ionize at these parameters of the
field. These results were shown previously on the FLI stability
plots in Fig.~\ref{fig:fig7}. Close inspection of the top and
central panels shows that the probabilities for the higher polarizations
are shifted to the left with respect to the probabilities for the lower
polarizations. For example, the maximum for the $(a)$ probability from
the central panel happens at $F=0.13$ and the maximum for the
$(d)$ probability from the left panel occurs at $F=0.14$.

On the bottom panel of Fig.~\ref{fig:fig8} the ionization probabilities
are shown for polarizations of the field close to the CP limit.
The fact that the ionization probabilities stay close to each other for
polarizations $\alpha\in[0.8, 1]$ demonstrates the lack of
significant variations of dynamics in the vicinity of the Lagrange
maximum. The probabilities change from the $80\%$ for $F=0.12$ up
to $100\%$ for $F=0.18$. Moreover, the ionization curves follow
almost monotonic growth with the increase of the amplitude of the
field. For the CP limit all the trajectories around the Lagrange
maximum become chaotic and ionize for $F=0.15$. This is the lowest
ionization bound observed.

\subsection{Phase space dynamics below the saddle point of zero--velocity surface}
\label{sec:4.3}

In this section we discuss the FLI stability results for the
ensemble of states with initial energy below the ZVS saddle point.
In addition, non-zero constant magnetic field was applied
perpendicular to the plane of polarization.

\begin{figure}
 \includegraphics[width=4.cm,angle=-90]{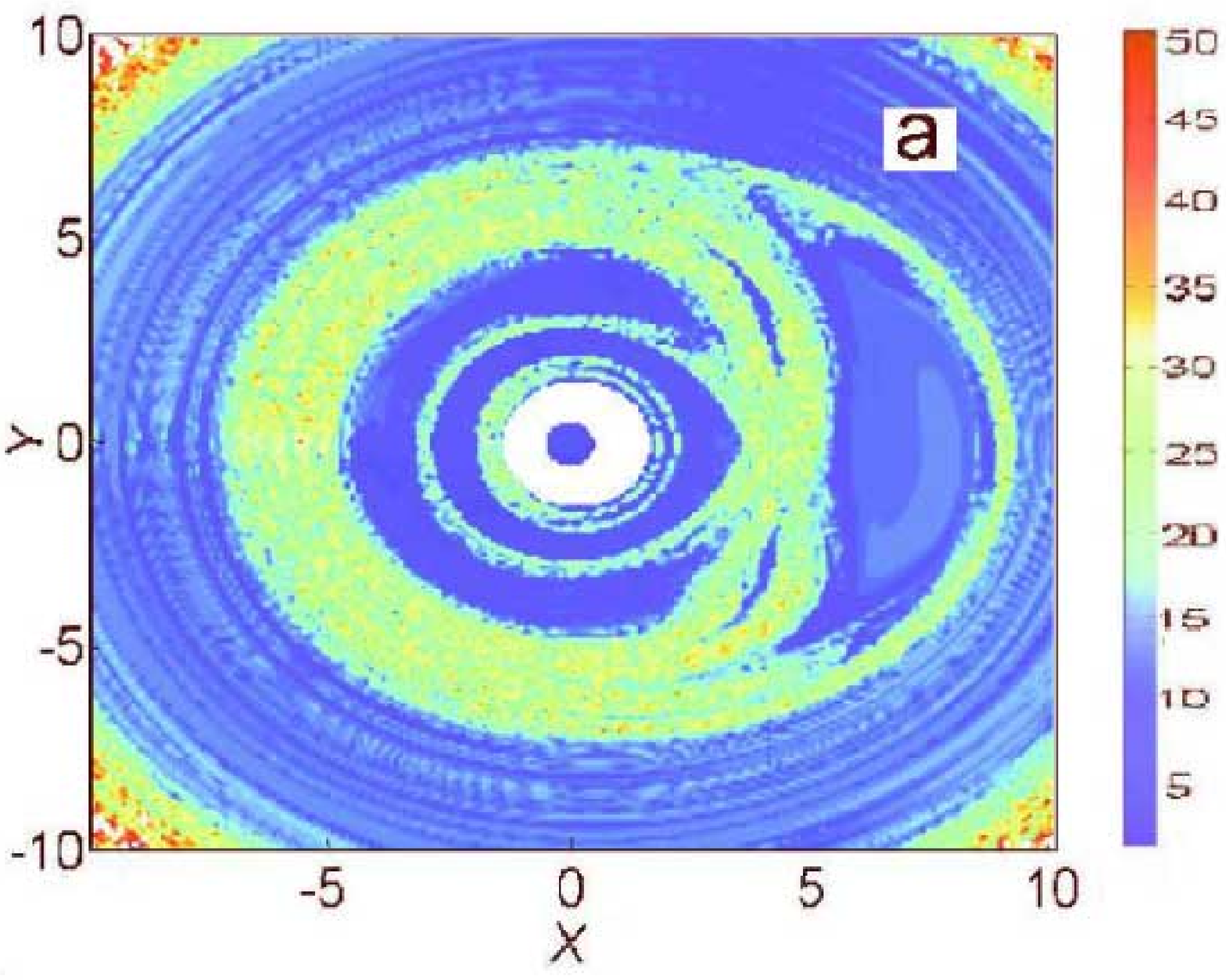}
 \includegraphics[width=4.cm,angle=-90]{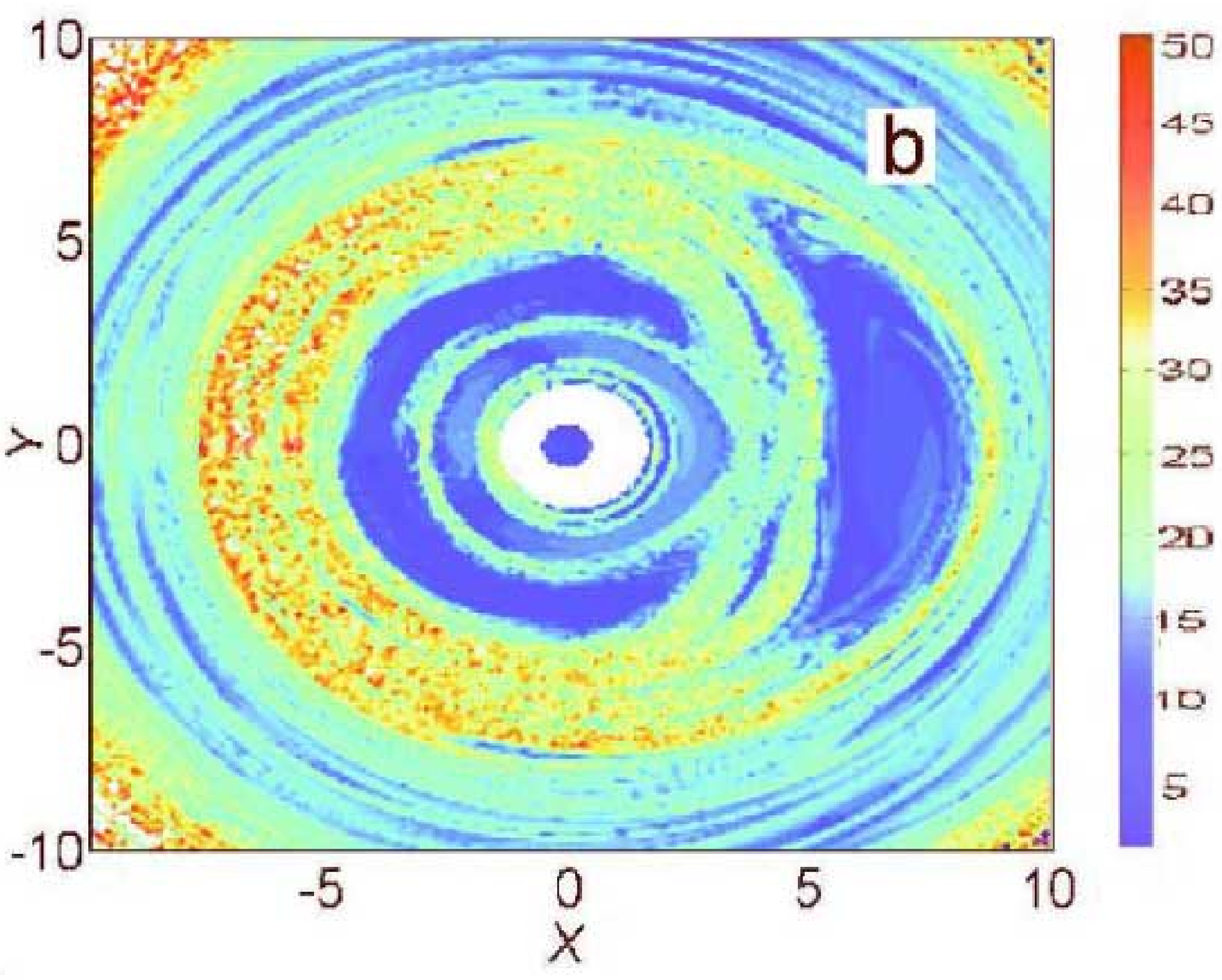}
 \includegraphics[width=4.cm,angle=-90]{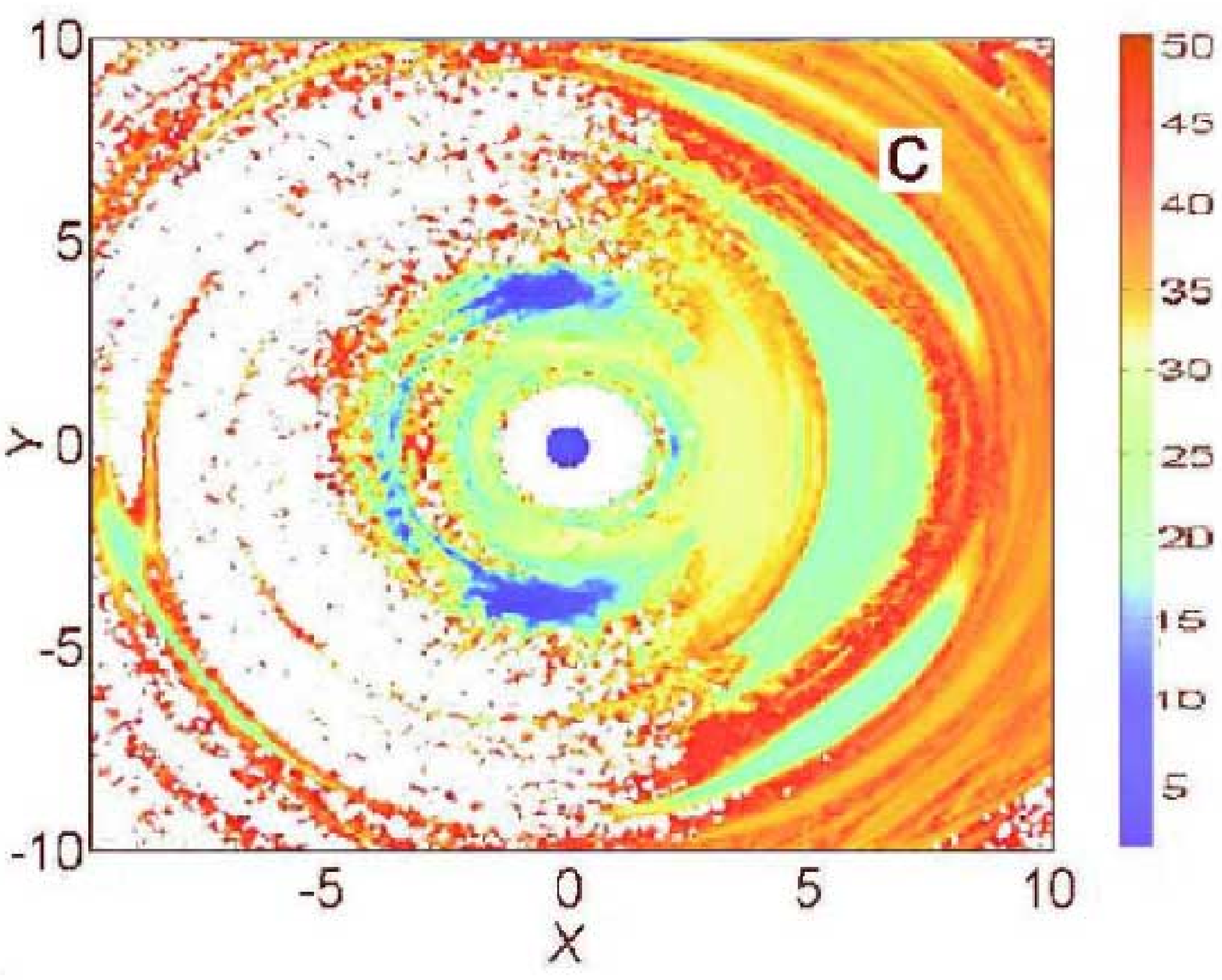}
 \includegraphics[width=4.cm,angle=-90]{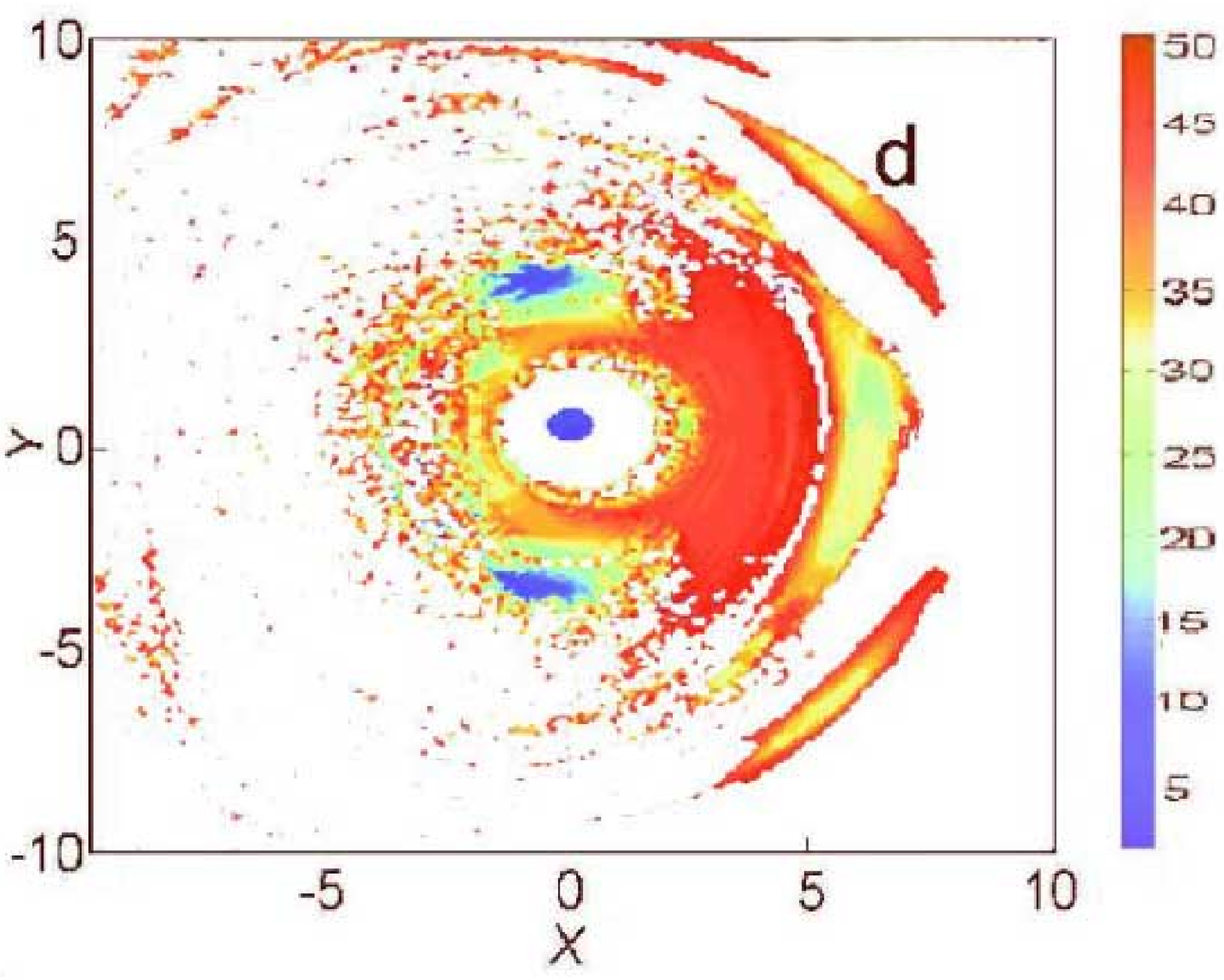}
 \caption {\label{fig:fig9} (Color online) FLI stability plots for hydrogen in an EP
 microwave field. $F=0.117,B=0.2,K=-2$. The polarization of the field
 is : a) 1, b) 0.8, c) 0.5, and d) 0.3.
 The color scale is assigned according to the maximum values of the FLI attained
 over the integration interval $t=100$.}
\end{figure}

 Previously it has been shown that the application of a magnetic field
perpendicular to the polarization plane leads to stabilization of
some of the phase space invariant tori for
hydrogen in a CP field \cite{Lee97,chand02}. In turn, the invariant
tori create barriers to the diffusion of chaotic trajectories
within the phase space. In this section we perform FLI stability
analysis for the hydrogen in EP microwave field and non--zero
magnetic field. As a starting point, the FLI stability analysis is
applied to hydrogen in CP field for the same parameters of the
electric and magnetic fields. The FLI calculations are carried out
on a grid of points from the $x-y$ plane for magnetic field
$B=0.2$ and initial energy $K_0=-2$ (the saddle point value of the
ZVS is $K_{sad}=-1.7022$). The resulting FLI stability plots are
shown in Fig.~\ref{fig:fig9}.

On panel $(a)$ the structure of the phase space is shown for the
CP field. Note the dark (blue in color version) island of
stability around the nucleus. The same stable island was observed
in Figs.~\ref{fig:fig3}, \ref{fig:fig4} and \ref{fig:fig7}. It is
located around the center of the ZVS. The island is surrounded by
the classically inaccessible region. The phase space outside of
the region is foliated with regular and chaotic motions. Chaotic
dynamics is associated with  high values of the FLI and color
coded with light colors (yellow and red in color version). The
large resonant zone is located on the right from the center of the
FLI plot. The phase space appears to be mostly regular, except for
thin chaotic layers winding around the large resonant zones. It is
easy to notice that the structure is symmetric with respect to the
$y$ axis. On panel $(b)$ the phase space dynamics is pictured for
the polarization $\alpha=0.8$. The structure resembles the one
shown on panel $(a)$: the FLI plot shows the same resonant
structures and chaotic layers as those observed on
panel $(a)$. However, the overall dynamics appear more chaotic
than for the CP field case. The stability results for the
intermediate polarizations $\alpha=0.5$ and $\alpha=0.3$ are shown
on the panels $(c)$ and $(d)$ correspondingly. The phase space
dynamics on both panels appear to be much more chaotic than the
dynamics for the CP limit. Although the large part of the phase
space on contour plots $(c)$ and $(d)$ corresponds to strongly
chaotic motions, several stable islands can be distinguished. For
example, two small islands of stability are located symmetrically
with respect to the $y$ axis.

Our main observation from the FLI stability results described in
this section is that the application of magnetic field leads to
the stabilization of the resonant structures within the phase
space of the system. The FLI plots reflect the onset of
stochasticity in the phase space of the system that occurs when
the field polarization approaches LP limit. The phase space
structure appears to be more regular for the field polarizations
close to the CP limit as opposed to the polarizations close to the
LP limit.

\subsection{Description of an ensemble of initial states and
comparison with the stability analysis results}
\label{sec:4.4}

Finally, in this section we introduce the classification of the
ensemble of orbits used for the FLI stability calculations in the
previous sections. In the  experiment \cite{beller97} the
ensemble of initial states are prepared spanning a narrow range of
the Keplerian energy $E=-\frac{1}{2n^2}$ and a variety of high
eccentricity orbits. As the field is turned on, the states with
the same action $n$ will appear in different regions of the phase
space having different values of energy and eccentricities
$\epsilon=\sqrt{1-l_z^2/n^2}$ ($l_z$ is angular momentum) in the
rotating frame. While the Hamiltonian dynamics is traditionally
studied by restricting the phase space to the constant energy
manifold, one would ideally prefer to relate numerical results
with the experiment and to study an ensemble of states with the
same Keplerian energy. However, most of the numerical simulations
based on the examining the structure of the Poincar\'{e} section
use trajectories with different values of the Keplerian energy
$K$. For the FLI stability analysis we use an ensemble of initial
states with equal initial energies $K(0)$ and different values of
initial action $n$ and eccentricity $\epsilon$. We choose these initial states to
compare the stability pictures given by the FLI method with the
Poincar\'{e} sections.

The interpretation of numerical results and comparison with experiments is much more complicated for the EP problem.
The presence of the time--dependent term in the expression of the
Hamiltonian in the rotating frame prevents the one-to-one relation
between the initial state (at time $t=0$) and the final state
(obtained after the turn--on of the field). Therefore, it is
instructive to relate the FLI stability results for the ensemble
of states analyzed with their initial eccentricities, actions and
angular momenta. The importance of the configuration of
initial states in the ionization mechanism was realized, e.g., in
Refs.~\cite{kappertz93,farrelly95,brunello96}. We found that our results
agree in their key features with the conclusion stressed out in
Ref.~\cite{brunello96}: the fate of the initial states subjected
to the application of the field cannot be predicted from the
initial action values and eccentricities of such states. In
essence, the states with the same initial action and eccentricity
values can be moved by the application of the field to different
parts of the phase space. We illustrate these conclusions by
comparing the FLI stability results given in Secs.~\ref{sec:4.1}
and \ref{sec:4.3} with the configuration of initial states
involved in the FLI computations.
\begin{figure}
 \begin{center}
 \includegraphics[width=4.cm]{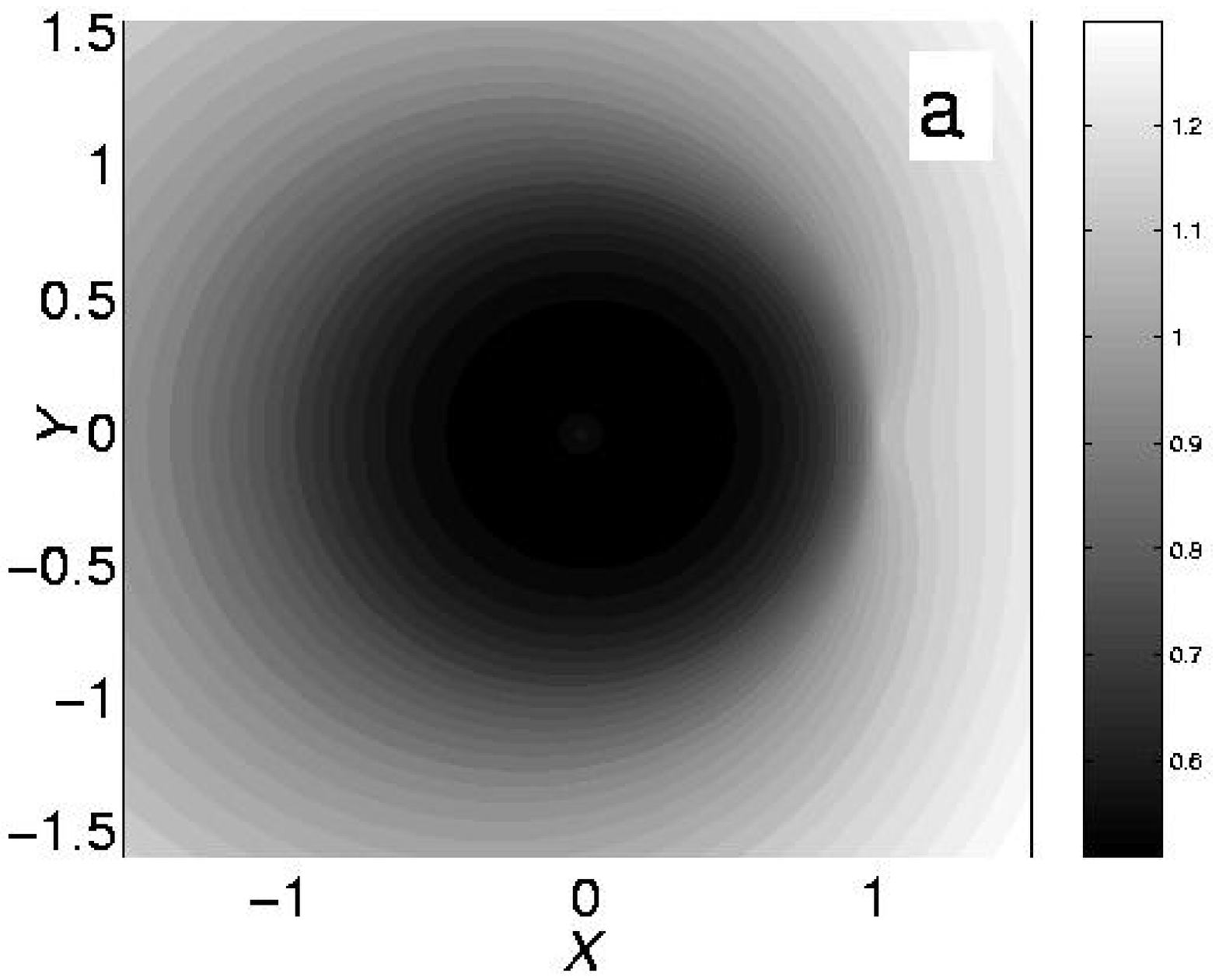}
 \includegraphics[width=4.cm]{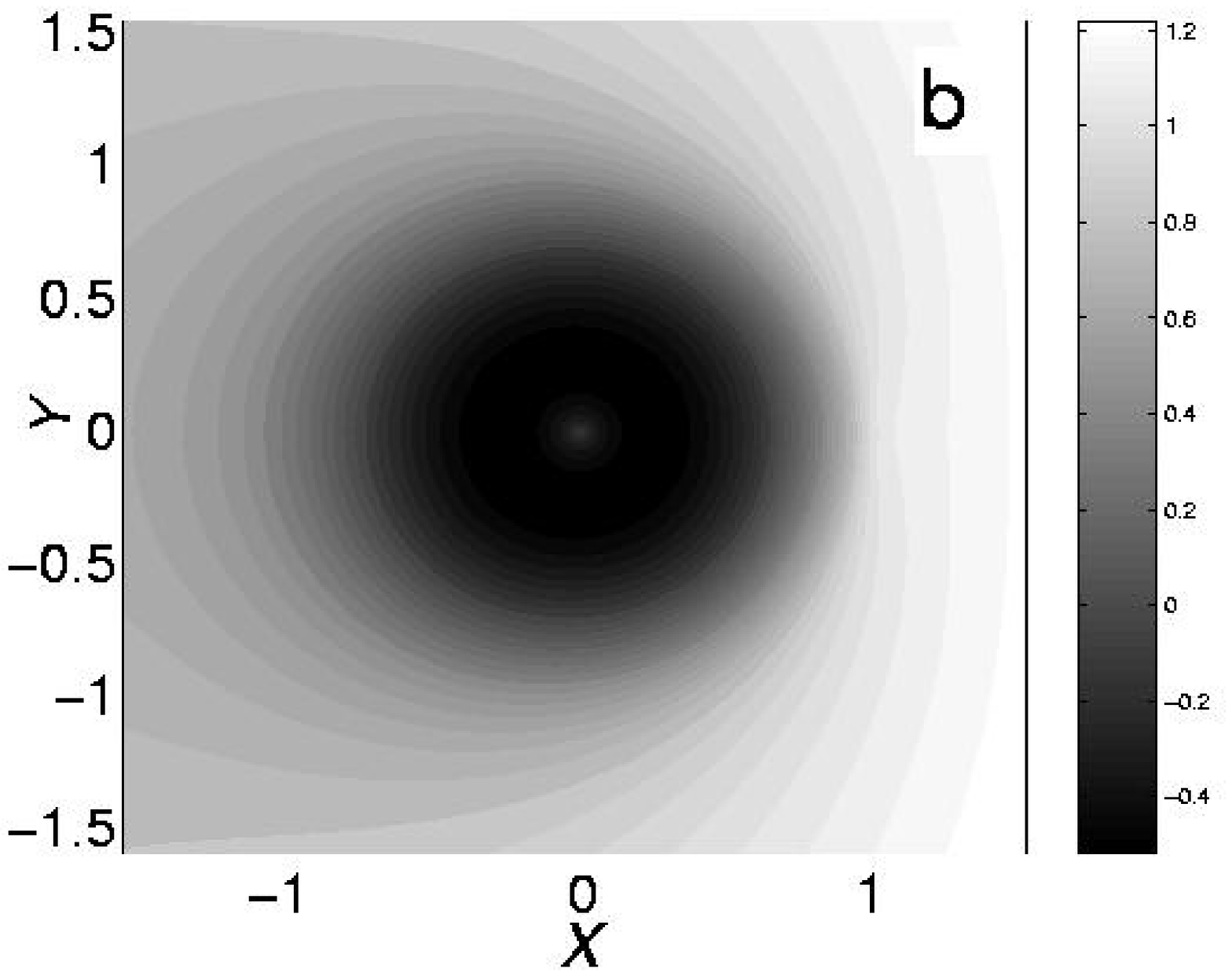}
 \includegraphics[width=4.cm]{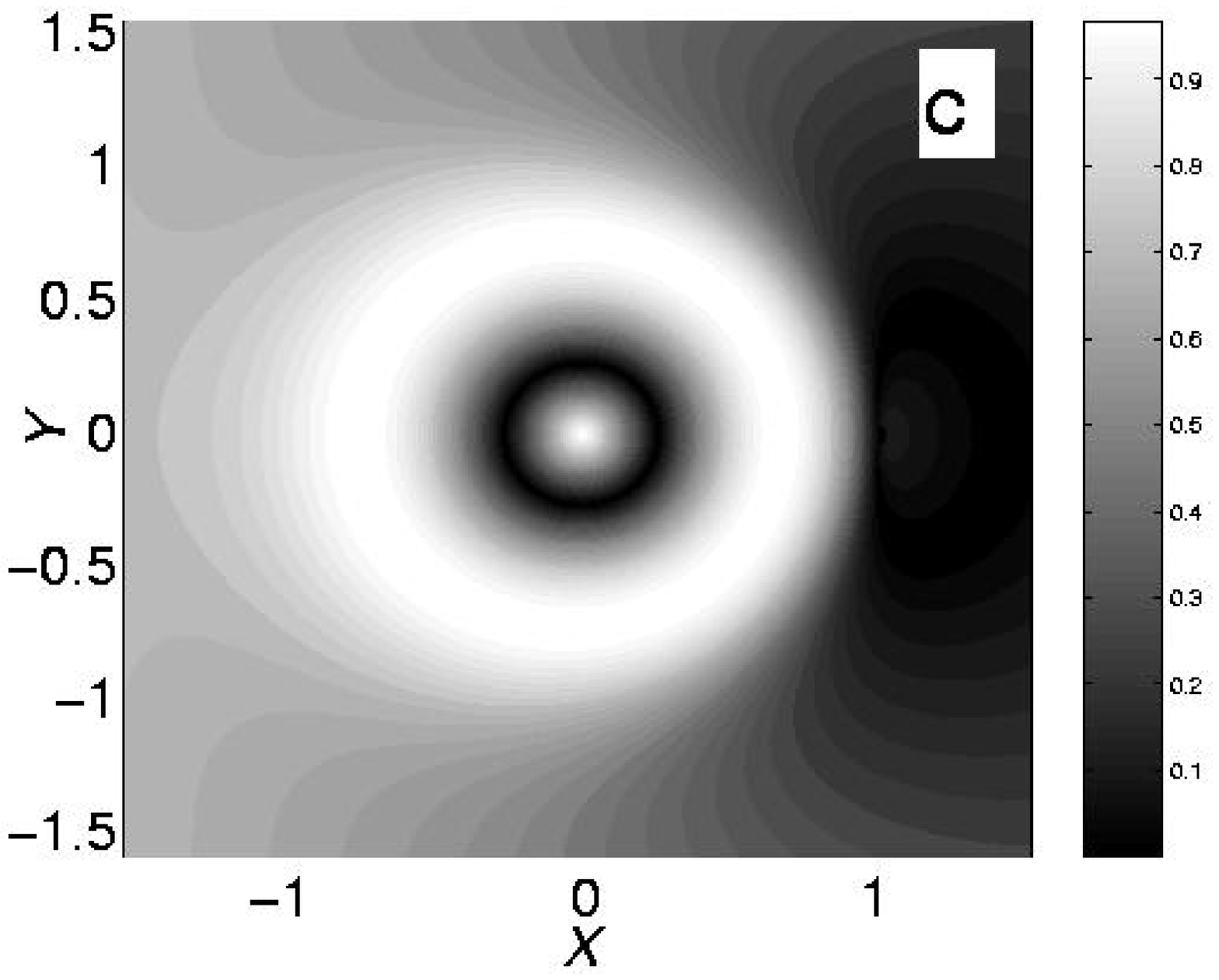}
 \end{center}
 \caption {\label{fig:fig10} The action variable $n$,
 angular momentum $l_z$, and eccentricity
 $\epsilon$ of the initial states of hydrogen in the EP microwave field
 are shown on panels  $(a), (b),$ and $(c)$ correspondingly. The parameters are $F=0.117, K_{max}=-1.3807$, and $B=0$.}
\end{figure}

First, we analyze the ensemble of states with initial energy equal
to the maximum of the ZVS. In Fig.~\ref{fig:fig10} the action
variables $n$, angular momenta $l_z$ and eccentricities $\epsilon$
are computed for the ensemble of initial states used for the FLI
stability analysis in Sec.~\ref{sec:4.1}. The case of zero
magnetic field is considered. The scaled amplitude of the electric
field is $F=0.117$, polarization is $\alpha=1$, and the initial
energy value is equal to the maximum $K_{max}=-1.3807$ of the ZVS
defined in Eq.~$(9)$. On panel $(b)$ the values of initial angular
momentum $l_z$ are shown. It can be seen from
Fig.~\ref{fig:fig10}$(b)$ and FLI plots in Fig.~\ref{fig:fig7}
that the states inside the central stable island correspond to
negative angular momentum states. The rest of the trajectories in
the FLI plots in Fig.~\ref{fig:fig7} corresponds to positive
angular momentum initial states. Another important observation is
that the central stable island is associated with bounded states
of different eccentricities. From the stability results given in
Sec.~\ref{sec:4.1} these states remain bounded for the range of
the scaled field amplitude $[0.117, 0.2]$ and for different
polarizations of the field. From Fig.~\ref{fig:fig10}$(c)$ and FLI
plots in Fig.~\ref{fig:fig7} it is apparent that the initial
states around the maximum of ZVS are the low eccentricity states.
In fact, these are the states that determine the low ionization
threshold computed in Sec.~\ref{sec:4.2}. Summarizing data given
in Figs.~\ref{fig:fig7} and \ref{fig:fig10} we argue that the
ultimate fate of initially low eccentricity and positive angular
momentum states is to be moved by the driving field to the chaotic
region of the phase space and undergo fast ionization. At the same
time, the initial states with negative value of angular momentum
located around the center of the ZVS remain bounded.
\begin{figure}
 \begin{center}
 \includegraphics[width=4.cm]{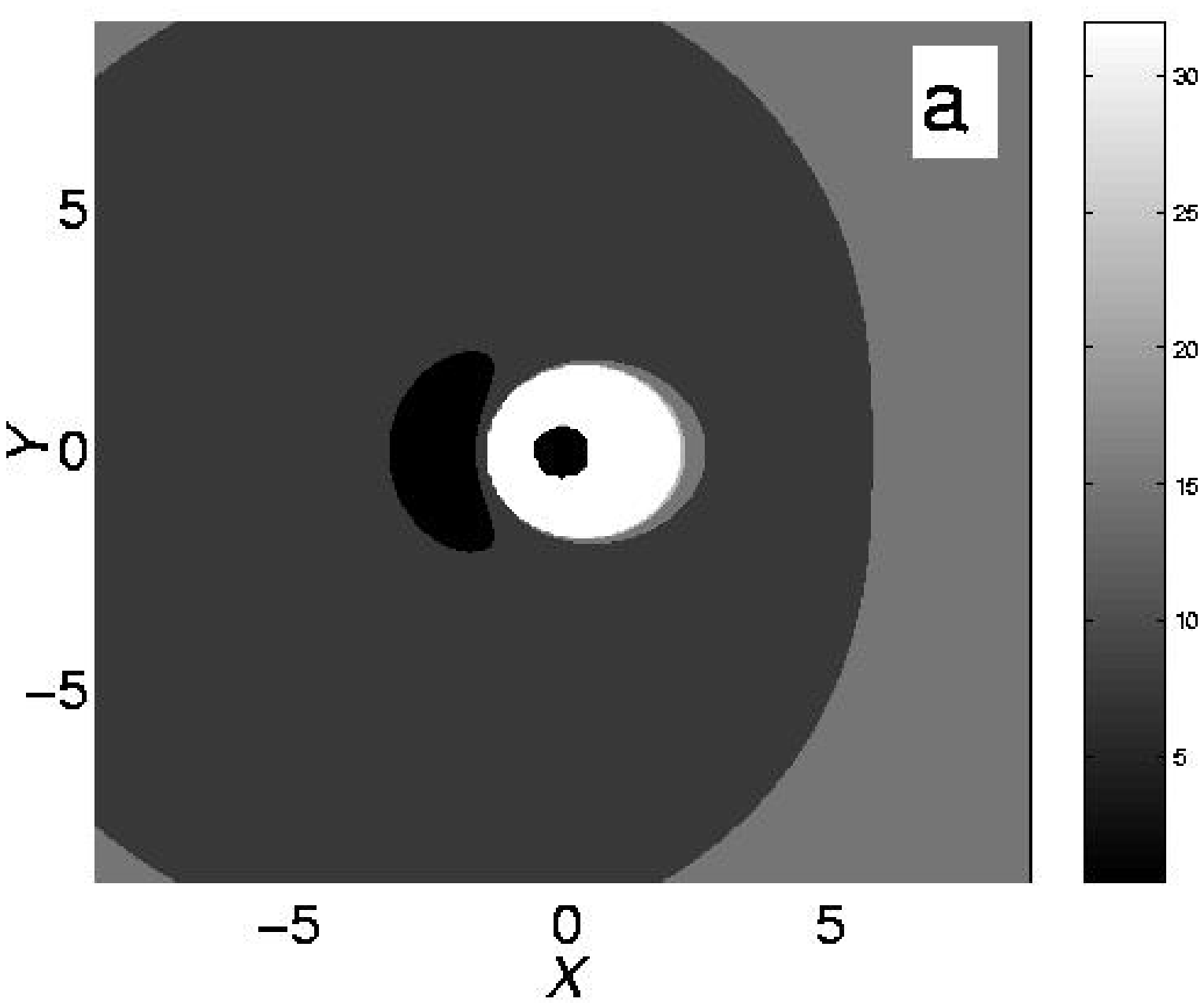}
 \includegraphics[width=4.cm]{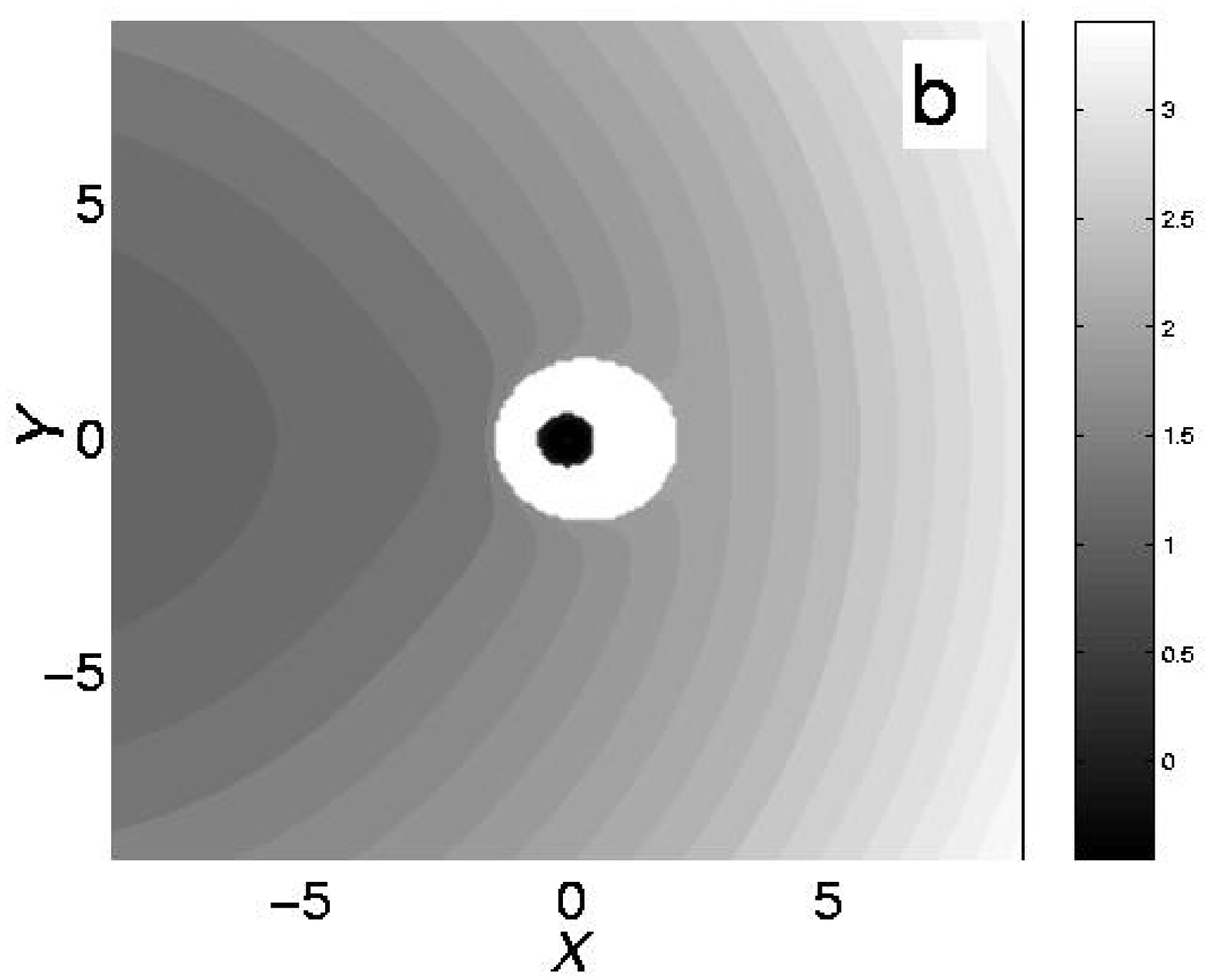}
 \includegraphics[width=4.cm]{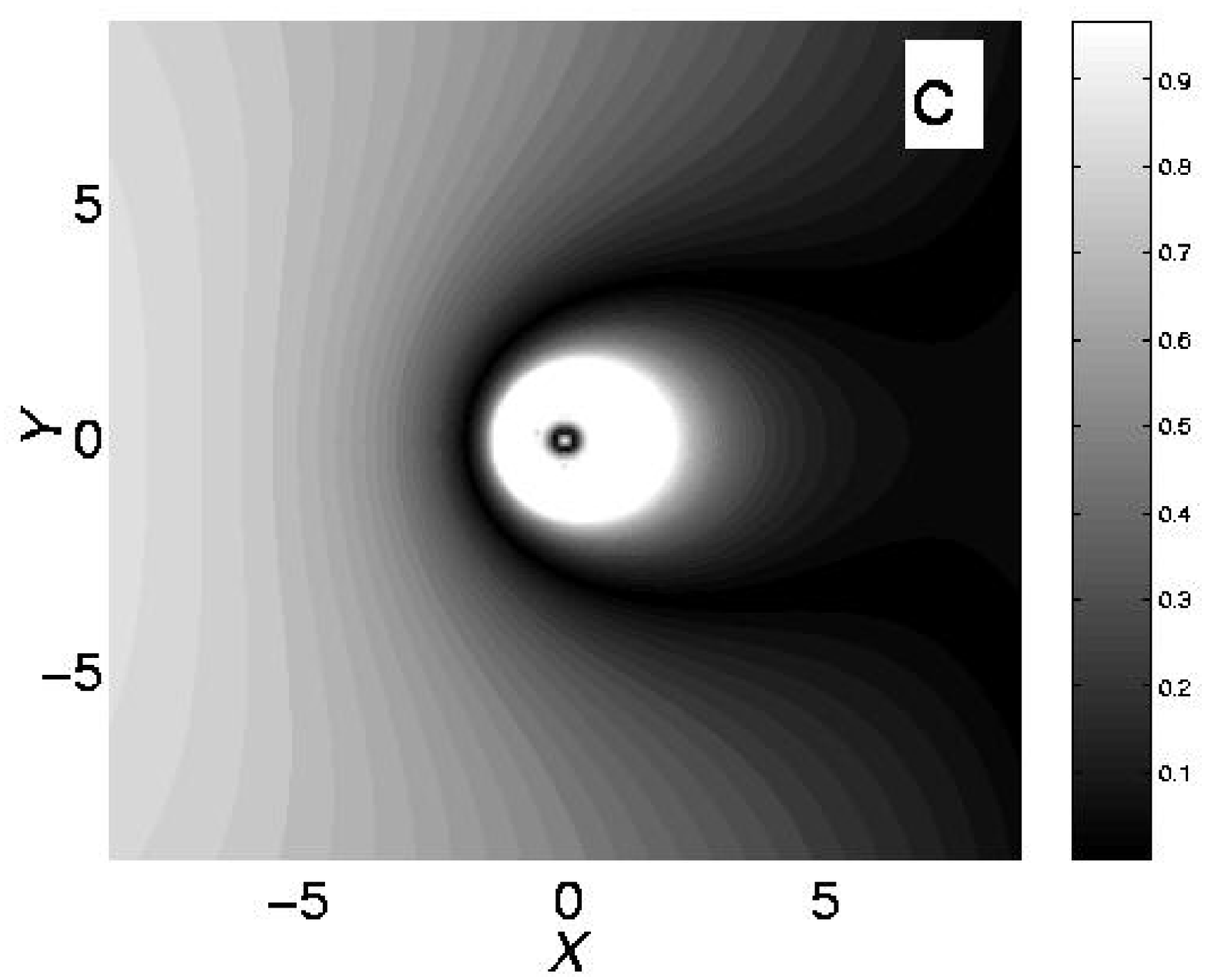}
 \end{center}
 \caption {\label{fig:fig11}  The action variable $n$,
 angular momentum $l_z$, and eccentricity
 $\epsilon$ of the initial states of hydrogen in the EP microwave field
 are shown on panels $(a), (b),$ and $(c)$ correspondingly. The
 parameters are
 $F=0.117, K=-2$, and $B=0.2$. The initial energy of the states $K$ is taken below the saddle point energy $K_{sad}=-1.7022$.}
\end{figure}

Next, we classify the initial states with energy below the saddle
point of the ZVS. In Fig.~\ref{fig:fig11} the action, angular
momentum and eccentricity together with the FLI plot are
presented for the initial states of hydrogen subject to EP field
$F=0.117$ with polarization $\alpha=0.9$ and magnetic field
$B=0.2$. The initial energy $K(0)=-2$ is chosen below the saddle
point $K_{sad}=-1.7022$ of the ZVS. The main conclusion from the
data given in Figs.~\ref{fig:fig9} and ~\ref{fig:fig11} is that
the initial states with low eccentricity tend to be more stable
than the states with high eccentricity values. For example, from
Fig.~\ref{fig:fig9}$(d)$ one observes three small stable islands
with low values of the FLI: one island is around the center of ZVS
and the other two islands are symmetric with respect to the
$y$--axis. Two symmetric islands correspond to low eccentricity
states in Fig.~\ref{fig:fig11}$(c)$.

The comparison of data presented in this section with the
stability results given by FLI analysis show that, although the fate of initial states can be determined from their
eccentricity and angular momentum values in
some cases, in general, this
information is not sufficient for predicting the dynamics of the
states after the turn--on of the field. Instead, the analysis of
the character of the states should be performed in conjunction
with the stability analysis.

\section{Conclusions}
\label{sec5}

In this article, we provide a qualitative description of the classical phase space
and its relevance to the ionization of hydrogen atom in a strong
EP microwave field. Using the FLI stability analysis, the
complex multidimensional dynamics is shown to depend sensitively
on the changes of parameters including polarization, amplitude of
the electric and magnetic fields and the initial state ensemble.
The FLI stability analysis allows us to picture complex phase
space structures, such as important resonant and chaotic zones.

We map out the FLI values for each trajectory originating on the
zero--velocity subspace for the ensemble of orbits with initial
energy at the maximum of the ZVS. Our FLI stability computations
for the LP and CP fields reveal two main resonant
structures: a small stable island around the Lagrange maximum of the
ZVS and a large stable island around the center of the ZVS. The
small stable island corresponds to the initial states with low
eccentricity values and positive angular momenta. In contrast,
the initial states from large stable island are the negative angular
momentum states with different eccentricity values. These states
remain bounded and their dynamics is almost unaffected by the
application of the microwave field. The remaining part of phase
space is attributed to chaotic motions that ionize quickly over
the interval of time considered in the computations. The main
feature of FLI stability results is that the ionization dynamics
is determined by the orbits with low eccentricity and positive
momentum states located around the maximum of the ZVS. These states
are the first to become chaotic and ionize for a given strength of
the field. Similar observations hold for the EP field case:
The FLI plots reflect stable structures similar to those observed
for the LP and CP problems. The main distinction from the LP and CP
field results is the behavior of stable island around the Lagrange
maximum. It changes non--monotonically with the increase of the
amplitude of the electric field due to break--up and stabilization
of resonant torus structures within the island.

Our main conclusion is that the nonlinear stability of the
Lagrange maximum is determined by the parameters of the field,
such as the amplitude of electric and magnetic field as well as
the field polarization. The ionization probabilities versus the amplitude
of the electric field reflect the changes in the local dynamics
around the Lagrange maximum. They manifest non--monotonic growth
for intermediate ellipticities of the field. This behavior agrees
with the experimentally observed ionization yields in Ref.~
\cite{beller97}.

 The FLI stability analysis was also carried out for the ensemble of states
 with initial energy below the saddle value of the ZVS. In this case the FLI plots
 reflect the onset of stochasticity in phase space that takes place
 as the field polarization changes from the circular
to the linear limit. The FLI simulations reveal the detailed
structure of phase space that is more regular for the field
polarizations close to the circular limit as opposed to the
polarizations close to the linear limit.

 The application of the method can be potentially important for
the control of ionization of Rydberg states in recent
experiments ~\cite{sirko02}. The main advantage of this technique
over other available analytical and numerical methods is its
independence from the dimensionality of the system. Thus, the stability
analysis can be carried out for any subspace of initial conditions
in the full--dimensional phase space of the system.

The method can be useful for
estimating the size of a stable zone around the equilibria for the wave packet calculations and
stabilization of quantum wave packets at the Lagrange equilibria
points discussed in Ref.~\cite{Lee97}. We
expect that by varying polarization and amplitude of electric and
magnetic fields, one can achieve the stabilization of a set of
resonances around the Lagrange maximum that can possibly lead to
the stabilization of the quantum wave packets in the EP field
system.

\begin{acknowledgments}
This research was supported by the US National Science Foundation.
CC acknowledges support from Euratom-CEA (contract
EUR~344-88-1~FUA~F).
\end{acknowledgments}

\end{document}